\documentclass[letter,11pt]{article}
\usepackage[T1]{fontenc} 
\usepackage{jheppub}
\usepackage{multirow}
\usepackage{amsmath}
\usepackage{slashed}
\usepackage{float}
\usepackage{cancel}

\usepackage{booktabs}

\newcommand\mf{{\sc\small MadFKS}}
\newcommand\ml{{\sc\small MadLoop}}
\newcommand\ct{{\sc\small CutTools}}
\newcommand\nin{{\sc\small Ninja}}
\newcommand\coll{{\sc\small Collier}}

\def\alphas{\alpha_s}

\def\TO{{\rightarrow}}
\def\ord{\mathcal{O}}
\def\alphas{\alpha_s}
\def\LO{\rm LO}
\def\NLO{\rm NLO}
\def\ljet{j_l}
\def\bjet{j_{b}}
\def\mttrue{m(t)}
\def\mtrec{m(t^{\rm rec})}
\def\ptrec{p_{T}(t^{\rm rec})}
\def\mebj{m(e^+ \bjet)}
\def\coselj{\cos(\theta^t_{e^+\ljet})}
\def\cosbjlj{\cos(\theta^t_{\bjet\ljet})}
\def\met{\slashed{E}_T}
\def\SoB{(Single-Top)/($W$+jets)}

\newcommand{\MSbar}{{\rm \overline{MS}}}
\def\nameprocess{\nu_e J J J}

\title{
Precise predictions for single-top production: the impact of EW corrections and QCD shower on the $t$-channel signature}

\author[a]{Rikkert Frederix,}
\author[b]{Davide Pagani,}
\author[b]{Ioannis Tsinikos}

\affiliation[a]{Theoretical Particle Physics, Department of Astronomy and Theoretical Physics, Lund University, S\"olvegatan 14A, SE-223 62 Lund, Sweden}

\affiliation[b]{Technische Universit\"{a}t M\"{u}nchen, James-Franck-Str.~1, D-85748 Garching, Germany}

\note{Preprint numbers: LU-TP 19-29, TUM-HEP-1216/19}

\abstract{In this work we calculate and provide precise predictions for the
  signature that is typically exploited at the LHC for the measurement of
  $t$-channel single-top production: 1 lepton, 1 light jet, 1 $b$-jet, $\met$
  and no additional jets or leptons. We apply the cuts that define the
  fiducial region and we take into account all the contributions to this
  signature; not only those from resonant $t$-channel single-top production.
  On the one hand, we calculate the complete-NLO corrections, {\it i.e.}, all
  NLO effects of QCD and EW origin. On the other hand, we study in detail the
  impact of a QCD parton shower for the fiducial region we consider. We
  provide predictions in different approximations at the inclusive level and
  for several differential distributions. Our study demonstrates the relevance
  of both EW corrections and shower effects for obtaining precise and reliable
  theoretical predictions for the single-top-production fiducial region.  }

\begin{document}
\maketitle

\section{Introduction}
The electroweak production of top quarks in hadron-hadron collisions can be categorised in three main production channels, which all  involve a single top (anti)quark in the final state and a $Wtb$ interaction vertex. The three different channels can be distinguished according to the
virtuality of the $W$ boson appearing in the Feynman diagrams. If the $W$ is space-like, the category
is called $t$-channel, if the $W$ is instead time-like the category
is called  $s$-channel and in the case of a final-state (on-shell) $W$ boson we refer to  $tW$ associated production.

 At the Large Hadron Collider (LHC) the category with the
largest production rate is the $t$-channel, with about 225~pb at 13 TeV, of which
approximately 135~pb is coming from top production and the rest from anti-top
production. Within the SM the interest in single-top production is mainly motivated by
the possibility of directly extracting the value of $|V_{tb}|$ in the CKM
matrix element~\cite{Alwall:2006bx,Lacker:2012ek,Cao:2015doa,Alvarez:2017ybk, Aaboud:2019pkc}. Moreover, in numerous BSM scenarios, single-top
production provides a sensitive probe to New Physics effects~\cite{Tait:2000sh,Atwood:2000tu,Drueke:2014pla,Aguilar-Saavedra:2017nik,Arhrib:2016vts,Jueid:2018wnj,Arhrib:2018bxc,Arhrib:2019tkr,Cao:2007ea,Zhang:2016omx,deBeurs:2018pvs}, possibly
parametrised through dimension-6 operators~\cite{Cao:2007ea, Zhang:2016omx, deBeurs:2018pvs, Hartland:2019bjb, Neumann:2019kvk}. Indeed,  involving the heaviest of the SM particles and EW interactions, this class of processes is of paramount interest for physics beyond-the-SM (BSM).

The total cross section for $t$-channel single-top production was first
measured at the Tevatron about 10 years ago~\cite{Abazov:2009ii, Aaltonen:2009jj}, albeit with large
uncertainties. The high-precision data from the LHC allowed not only for a
well-established measurement of the total production cross sections, but also
for a precise determination of differential distributions at the centre-of-mass
energies of 7, 8, and 13~TeV~\cite{ Chatrchyan:2011vp, Aad:2012ux, Chatrchyan:2012ep,  Khachatryan:2014iya,  Aad:2014fwa, Aaboud:2016ymp, Aaboud:2017pdi,  Sirunyan:2018rlu, CMS-PAS-TOP-17-023}. 

The top quarks decay (predominantly) to a $W$-boson and a $b$-quark. In order to be
able to tag the $t$-channel single-top production process experimentally, only leptonic
decays of $W$ bosons are considered. With hadronic $W$ decays  the signature would be
indistinguishable from the multi-jet background, which has a cross section
that is many orders of magnitude larger than the $t$-channel signature. Hence,
the typical signature that is analysed involves one $b$-jet from the top decay, missing
transverse-energy and one high-$p_T$ lepton from the decay of the $W$-boson emerging from the top-quark decay, and one
additional light jet, which is associated to the light parton present in the LO
$t$-channel matrix elements.

On the theoretical front, predictions for $t$-channel single-top quark production have
been known at the NLO accuracy in QCD for quite some time~\cite{Harris:2002md,Campbell:2004ch,Cao:2005pq,Molbitz:2019uqz} in the five
flavour scheme (5FS), and for about ten years in the four-flavour
scheme (4FS)~\cite{Campbell:2009ss,Campbell:2009gj,Campbell:2012uf}. Much more recently, also the NNLO corrections have been
computed~\cite{Brucherseifer:2014ama}, also including the decay of the top quark in the narrow
width approximation up to the same level of accuracy, but neglecting some
interference contributions in the NNLO corrections \cite{Berger:2016oht,Berger:2017zof}. The NLO corrections
in the electroweak coupling have been computed~\cite{Beccaria:2008av,Bardin:2010mz,Frederix:2018nkq} and are found to be
small for the inclusive production, but can be significantly enhanced in
certain regions of phase space. Going beyond the narrow-width approximation
for the top quarks, off-shell effects have been studied at fixed order
perturbation theory in refs.~\cite{Falgari:2010sf, Falgari:2011qa, Papanastasiou:2013dta}.

Beyond fixed-order perturbation theory, the all-order analytic resummation of
threshold and transverse momentum logarithms have been presented in
refs.~\cite{Wang:2010ue,Kidonakis:2011wy} and ref.~\cite{Cao:2018ntd}, respectively. Monte Carlo simulations, in which
the NLO single-top production processes have been matched to a parton shower
are available in the MC@NLO~\cite{Frixione:2005vw,Frixione:2008yi,Frederix:2012dh,Papanastasiou:2013dta,Frederix:2016rdc}, Sherpa~\cite{Bothmann:2017jfv} and POWHEG~\cite{Alioli:2009je, Frederix:2012dh, Jezo:2015aia}
frameworks. Within the latter framework, also the consistent inclusion of the
single-top plus one-jet NLO matrix elements has been considered through the
(extended) MINLO method~\cite{Carrazza:2018mix}.

The main focus of this paper is to re-visit single-top production in the
$t$-channel in a suitably defined fiducial region of phase space. The
motivation is as follows.  Although the categorisation of the three main
single-top production channels is well-defined at lowest order in perturbation
theory, it breaks down when including higher-order quantum
corrections. Indeed, interferences between the $t$-channel and the other
production modes are non-zero.  Also, from an experimental point of view, the virtuality of the
$W$-boson, which defines the categorisation, is not a direct
observable. This, formally, also breaks down the categorisation from an
experimental point of view. Therefore, in this work we do not include only
$t$-channel single-top production into our predictions and simulations. Rather,
we include all contributions that produce exactly an electron, a neutrino and exactly a
jet-pair, of which only one is $b$-tagged, and consider a fiducial region  that is
dominated by $t$-channel single-top production. We calculate all LO terms and
all the complete-NLO corrections, {\it i.e.}, all NLO effects of QCD and EW origin are taken into
account. Furthermore, we study in detail the impact of the parton shower on the predictions for the $t$-channel fiducial
region.

This paper is organised as follows. In Sec.~\ref{sec:frameworks} we describe the two different  frameworks used in this work: complete-NLO accuracy at fixed order, in Sec.~\ref{sec:fixressetup}, and NLO QCD corrections matched to shower effects, in Sec.~\ref{sec:showersetup}.
Then, we present numerical predictions at the inclusive and differential level for several phenomenologically relevant distributions in Sec.~\ref{sec:results}, for both the aforementioned approximations. Finally, in Sec.~\ref{sec:conclusions} we write our conclusions and we summarise our findings. Furthermore, in Appendix \ref{sec:comparison} we collect  comparisons  with previous results and among  different approximations.

\section{Calculational frameworks}
\label{sec:frameworks}

In this section we describe the two calculational frameworks on which the results presented in Sec.~\ref{sec:results} are based. 
Our main focus is providing precise predictions for $t$-channel single-top production with subsequent top-quark leptonic decays. Therefore, we consider the signature
\begin{equation} \label{signature}
e^+,~  1{\rm ~ light ~ jet},~ 1~ b{\rm-jet},~ \met {\rm ~and ~no ~additional ~jets}\, ,
\end{equation}
which is exploited in the measurements of $t$-channel single-top production at the LHC.  As already mentioned in the introduction, on the one hand, we calculate all the fixed-order NLO corrections of QCD and EW origin for the signature \eqref{signature}, {\it i.e.}, the complete-NLO prediction. On the other hand, we match a subset of the complete-NLO predictions to QCD shower effects. The calculation framework for the former approximation is discussed in Sec.~\ref{sec:fixressetup}, while for the latter in Sec.~\ref{sec:showersetup}. 

In our calculation, with both approximations,  we always apply cuts (they are explicitly defined in Sec.~\ref{sec:signal}) in order to select the fiducial region that has been considered by the ATLAS collaboration in the measurement of $t$-channel single-top production \cite{Aaboud:2017pdi}.  Other production processes contribute to the signature \eqref{signature}  as well and the fiducial region is defined in order to suppress them. 

It is important to note that, within our setup, the calculation for a signature similar to \eqref{signature} where a $\mu^+$ is present in the place of the $e^+$ is exactly the same.  Thus, the numbers given in this paper can be used also for that signature. On the other hand, in this work we do not consider the case where a $e^-$ (or $\mu^-$) is present in the place of the $e^+$, {\it i.e.}, the case of $t$-channel single-top antiquark production; we expect results to be qualitatively similar. 

\subsection{Fixed-order complete-NLO predictions}
\label{sec:fixressetup}

In this section we describe the calculation at fixed-order complete-NLO accuracy for the signature \eqref{signature}. First, in Sec.~\ref{sec:theory} we discuss the different resonances appearing in the different perturbative orders that enter the complete-NLO approximation and we introduce the notation used in this work.  Then, in Sec~\ref{sec:input} we specify the input parameters and finally in Sec.~\ref{sec:clustering} we describe the clustering procedure that we have adopted in order to ensure infra-red (IR) safety.  Numerical results are presented in Sec.~\ref{sec:fixres}.

The calculation that is described in this section, fixed-order complete-NLO predictions for the signature \eqref{signature}, has been performed via the latest version of {\sc\small MadGraph5\_aMC@NLO} \cite{Frederix:2018nkq}, which is public. In the  {\sc\small MadGraph5\_aMC@NLO} framework \cite{Alwall:2014hca} infrared singularities are dealt with via the FKS method~\cite{Frixione:1995ms,
Frixione:1997np}, which is automated in the module \mf~\cite{Frederix:2009yq,
Frederix:2016rdc}. The evaluation of one-loop amplitudes is performed by dynamically switching  among
different types of techniques for integral reduction, {\it i.e.}, the so called OPP method~\cite{Ossola:2006us},
Laurent-series expansion~\cite{Mastrolia:2012bu},
and  tensor integral reduction~\cite{Passarino:1978jh,Davydychev:1991va,Denner:2005nn}.
These techniques have been automated in the module \ml~\cite{Hirschi:2011pa}, which is employed for generating  the amplitudes. We recall that
\ml \, employs the codes \ct~\cite{Ossola:2007ax}, \nin~\cite{Peraro:2014cba,
Hirschi:2016mdz} and \coll~\cite{Denner:2016kdg}. Moreover, it includes an in-house
implementation of the {\sc OpenLoops} optimisation~\cite{Cascioli:2011va}.

\subsubsection{Structure of the calculation: underlying resonances and notation}
\label{sec:theory} 

\begin{table}[!t]
\begin{center}
\begin{tabular}{l  p{5cm}}
\toprule
Perturbative order & Resonant processes \\
\midrule
$\LO_1 ~(\alphas^2\alpha^2)$ & $W$ + 2 jets \\
$\LO_2 ~(\alphas\alpha^3)$ & - \\
$\LO_3 ~(\alpha^4)\phantom{\alpha^2}$ & single-top ($t$- and $s$-ch.), $WZ$ \\
\midrule
$\NLO_1 ~(\alphas^3\alpha^2)$ & $W$ + 2 jets\\
$\NLO_2 ~(\alpha^2\alpha^3)$ &  $W$ + 2 jets \\ 
$\NLO_3 ~(\alphas\alpha^4)$ & single-top ($t$- and $s$-ch.), $WZ$, $tW$, $\bar t W$ and $WW+b$-jet \\
$\NLO_4 ~(\alpha^5)\phantom{\alpha^2}$ & single-top ($t$- and $s$-ch.), $WZ$, $tW$, $\bar t W$ and $WW+b$-jet \\
\bottomrule
\end{tabular}
\end{center}
\caption{Intermediate resonances contributing to the various perturbative orders that enter the calculation.} 
\label{table:orders} 
\end{table}

The main process we are interested in is single-top production via $t$-channel with the top-quark decaying leptonically. In other words, $pp \TO t j$, where $j$ is a light jet and the top-quark is decaying into $t\TO e^+ \nu_e b$. In the 5FS, this process contributes at LO, which is of $\ord(\alpha^4)$, to the cross section for the signature \eqref{signature}. Due to the misidentification of $b$-jets as light jets, also $s$-channel single-top production contributes at the same order when the top quark decays leptonically. Similarly, $WZ$ production can contribute when the $Z$ boson decays into a $b \bar b$ pair. 
The $\ord(\alpha^4)$ contributions, however, are the formally smallest LO contributions in the expansion in powers of $\alphas$ and $\alpha$. Indeed, the signature \eqref{signature} receives also 
$\ord(\alpha_s^2 \alpha^2)$ contributions from tree-level diagrams:  $W$+jets with leptonic $W$ decays contributes to the $\ord(\alpha_s^2 \alpha^2)$. Thus, as already mentioned, single-top production is not the only production process contributing to this signature. Furthermore also non-resonant contributions are possible. 

\begin{figure}[!t]
\centering
\includegraphics[scale=0.4]{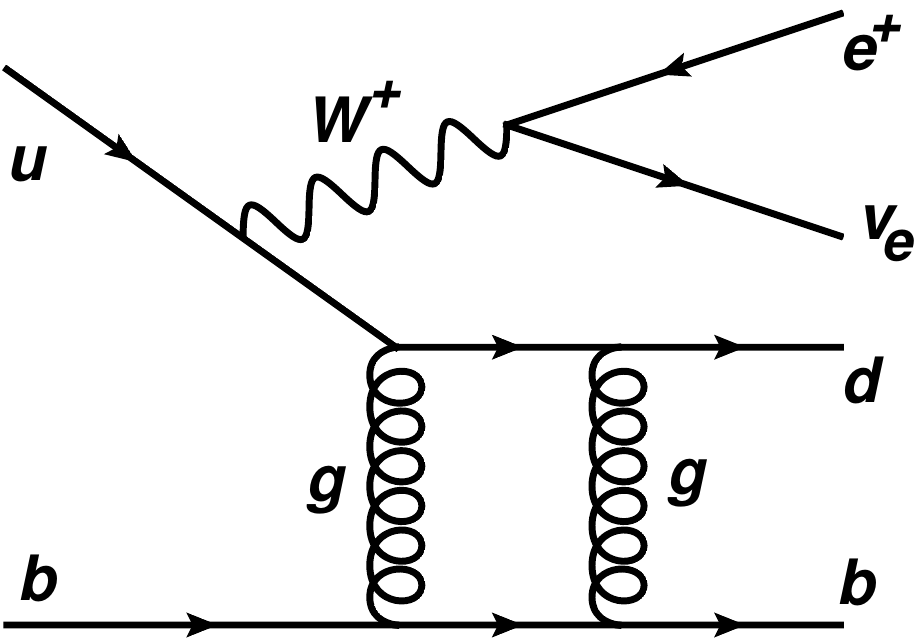}
\hspace{1cm}
\includegraphics[scale=0.4]{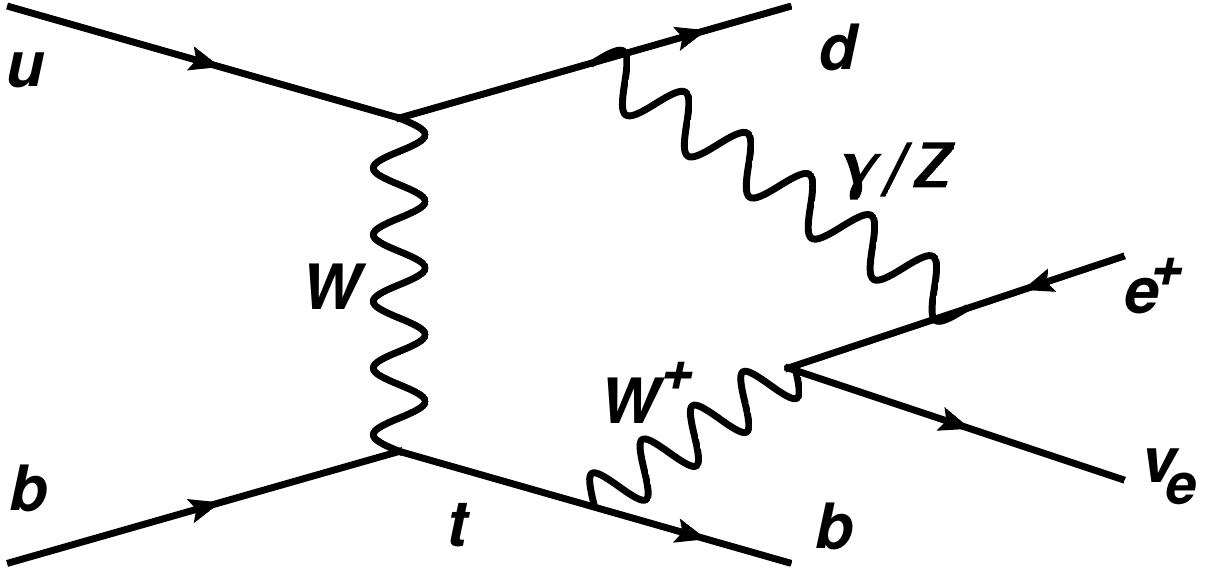}
\hspace{1cm}
\includegraphics[scale=0.375]{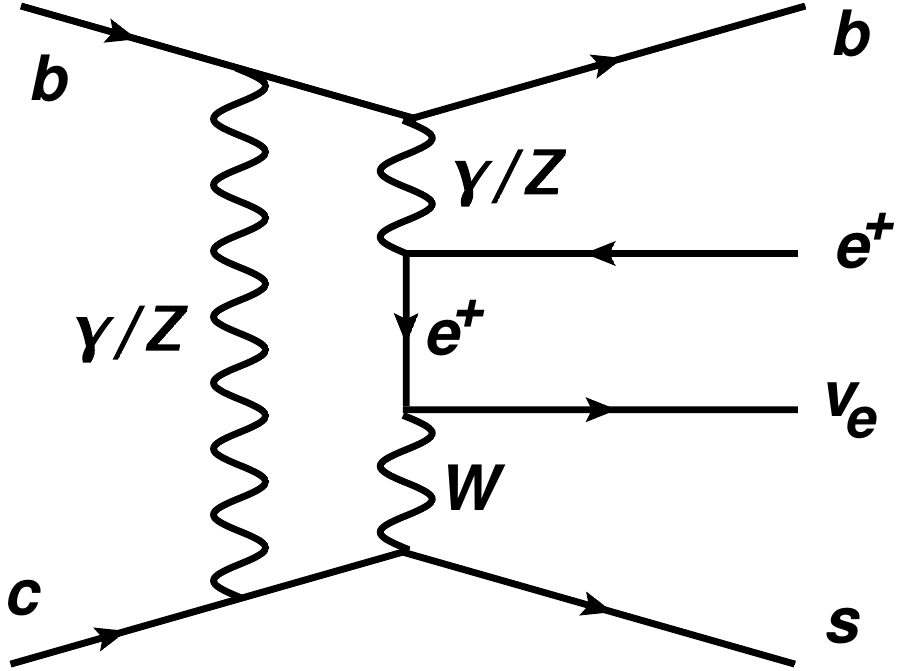}
\\
\vspace{0.75cm}
\includegraphics[scale=0.4]{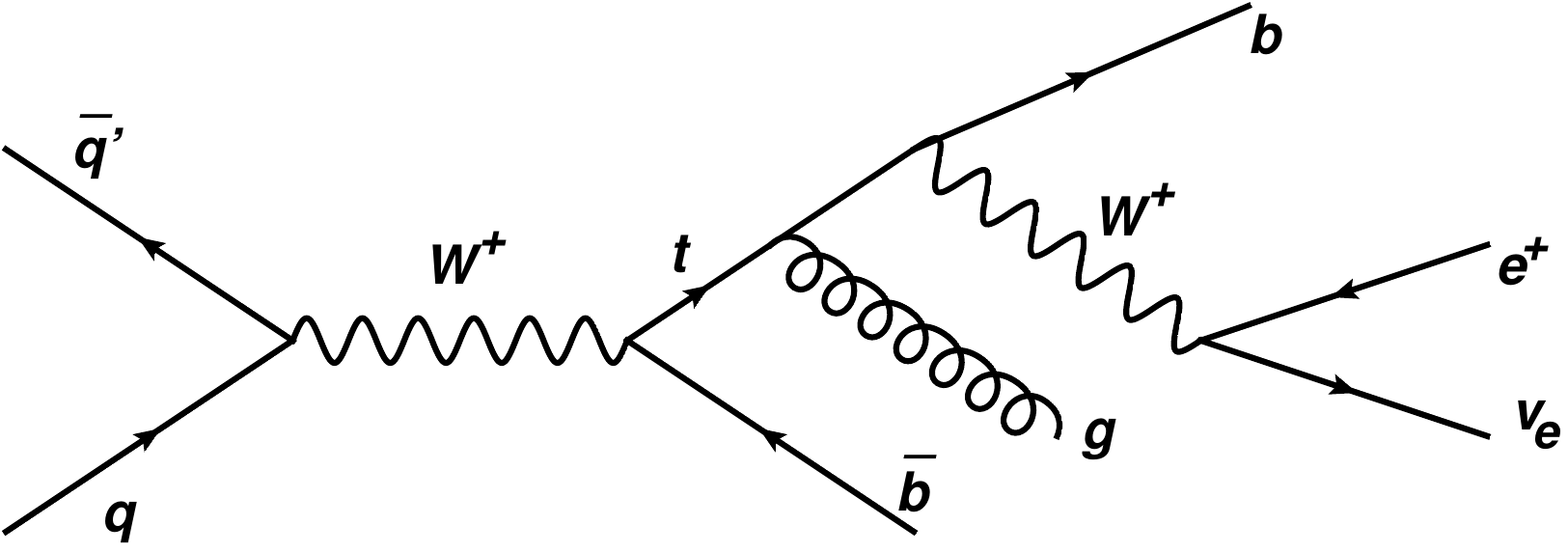}
\hspace{1cm}
\includegraphics[scale=0.375]{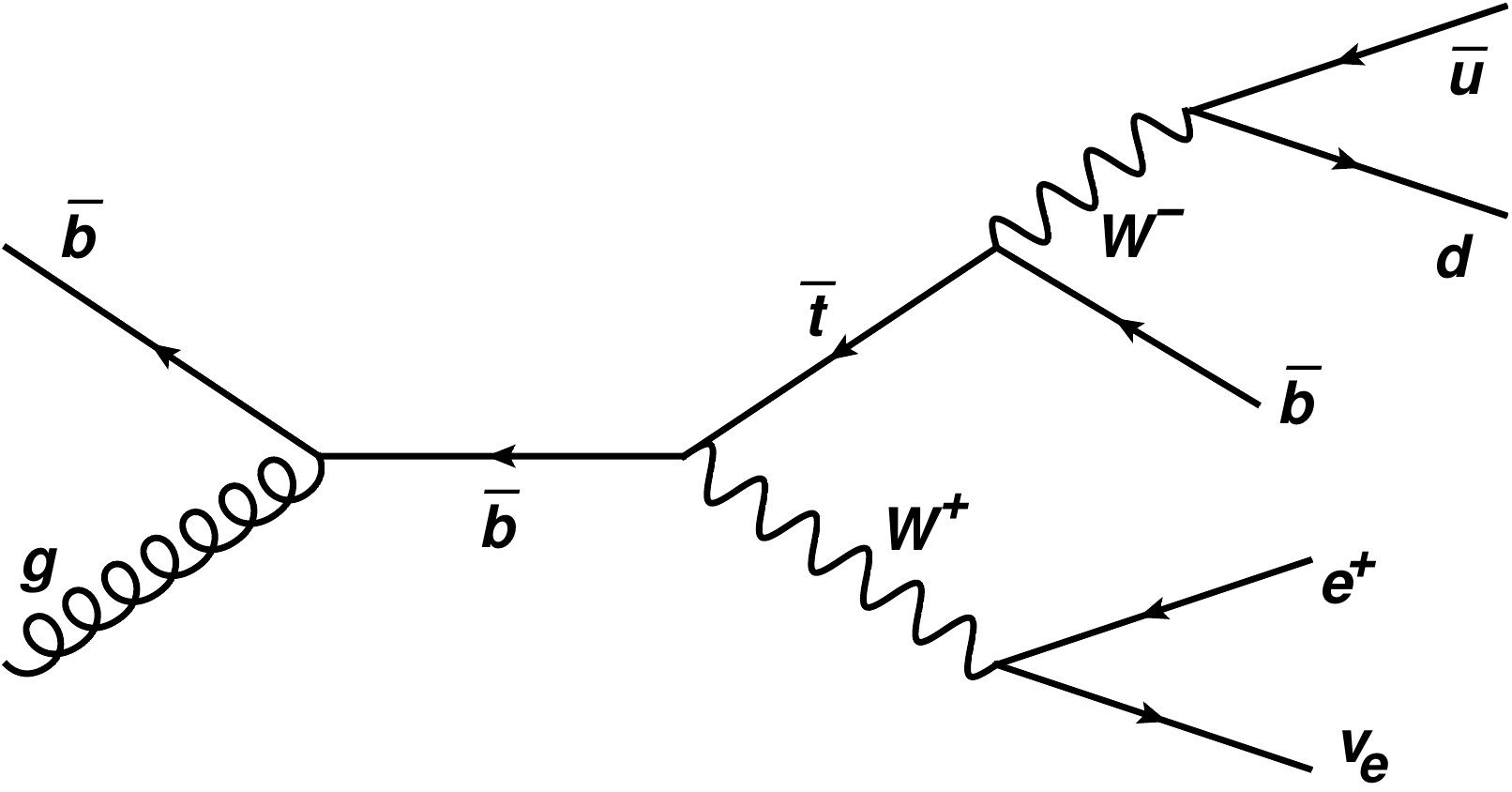}
\caption{Selection of Feynman diagrams contributing to the signature
  \eqref{signature}. The upper-left diagram contributes to $\NLO_1$
  and $\NLO_2$ ($W$+jets), the upper-central diagram to $\NLO_4$
  (single-top resonant), and the upper-right diagram also to $\NLO_4$
  (non-resonant). The lower-left diagram is a typical $s$-channel
  single-top production diagram, with an extra gluon, while the
  lower-right diagram can be considered $\bar{t}W^+$-associated
  production, both contributing to $\NLO_3$.}
\label{fig:diagrams}
\end{figure}

In this section we present the calculation of all the contributions to fixed-order complete-NLO predictions for the signature \eqref{signature}.
Following the notation already used in Refs.~\cite{Frixione:2014qaa, Frixione:2015zaa, Pagani:2016caq, Frederix:2016ost, Czakon:2017wor, Frederix:2017wme, Frederix:2018nkq, Broggio:2019ewu}, with complete-NLO predictions we denote all the one-loop and real emission corrections of QCD and EW origin. To this purpose we calculate all the $\ord(\alpha_s^m \alpha^{n+2})$ contributions  with $m,n>0$ and  $m+n=2,3$ to
\begin{equation} \label{syntax}
pp \to \nu_e J J J\, ,
\end{equation}
where $J$ is any particle that may potentially enter in a fully-democratic jet, {\it i.e.}, a jet that is obtained by clustering quarks (including $b$-quarks), gluons, photons and leptons.
As discussed in Refs.~\cite{Frederix:2016ost, Frederix:2018nkq}, this procedure is necessary in order to fully ensure IR safety when dealing with complete-NLO contributions and massless final state. In practice, given the presence of an electronic neutrino,\footnote{In our calculation lepton PDFs are safely set to zero \cite{Bertone:2015lqa}, so no initial-state leptons can be present.} all the possible final-states include a positron and two(three) massless particles.  

The different contributions to the total cross section can be denoted as:
\begin{align}
\Sigma^{\nameprocess}_{\LO}(\alpha_s,\alpha) &= \alpha_s^2 \alpha^2 \Sigma_{4,0}^{\nameprocess} + \alpha_s \alpha^3 \Sigma_{4,1}^{\nameprocess} + \alpha^4 \Sigma_{4,2}^{\nameprocess} \nonumber\\
 &\equiv \LO_1 + \LO_2 + \LO_3\, , \label{eq:blobLO} \\
 \Sigma^{\nameprocess}_{\NLO}(\alpha_s,\alpha) &= \alpha_s^3 \alpha^2 \Sigma_{5,0}^{\nameprocess} + \alpha_s^2 \alpha^3 \Sigma_{5,1}^{\nameprocess} + \alphas \alpha^4 \Sigma_{5,2}^{\nameprocess}+ \alpha^5 \Sigma_{5,3}^{\nameprocess} \nonumber\\
 &\equiv \NLO_1 + \NLO_2 + \NLO_3+  \NLO_4\, , \label{eq:blobNLO} 
\end{align}
Single-top production via $s$- and $t$-channel enters at $\LO_3$ and the corresponding NLO QCD and EW corrections are part of the $\NLO_3$ ({\it e.g.}~bottom-left diagram in Fig.~\ref{fig:diagrams}) and $\NLO_4$  ({\it e.g.}~top-central diagram in Fig.~\ref{fig:diagrams}), respectively. The same applies to $WZ$ production. $W$+jets contributes at LO to the $\LO_1$ and the corresponding NLO QCD and EW corrections are part of the $\NLO_1$ and $\NLO_2$ ({\it e.g.}~top-left diagram in Fig.~\ref{fig:diagrams}), respectively. Moreover, including NLO corrections, also $tW$-associated production can contribute to the signature \eqref{signature}. Indeed, due to the top-quark decay, two $W$ bosons are present: if one of them decays hadronically and the other one leptonically, LO contributions from  $tW$, $\bar tW$ ({\it e.g.}~bottom-right diagram in Fig.~\ref{fig:diagrams}) and $WW+b_{\text jet}$ production enter the $\NLO_3$ and $\NLO_4$. This pattern is summarised in Tab.~\ref{table:orders}. We remark that besides these production processes, all the off-shell and non-resonant effects ({\it e.g.}~top-right diagram in Fig.~\ref{fig:diagrams}) are exactly taken into account. 

In the following, in order to simplify the notation, we will also refer to the perturbative orders $\LO_3$, $\NLO_3$ and $\NLO_4$ as ``Single-Top'', while the remaining perturbative orders $\LO_1$, $\LO_2$, $\NLO_1$ and $\NLO_2$ will be also referred as $W$+jets. In particular,
\begin{align}
 &{\rm LO} = \LO_3\,, \nonumber\\
\text{Single-Top~~~}\longrightarrow ~~~&{\rm NLO~QCD} = \LO_3+ \NLO_3\,,  \nonumber\\
& {\rm NLO~QCD+EW} = \LO_3 + \NLO_3 + \NLO_4\,,\nonumber\\
{}\label{conversion}\\
 &{\rm LO} = \LO_1 (+\LO_2)\, ,\nonumber\\
W+\text{jets}~~~\longrightarrow ~~~&{\rm NLO~QCD} = \LO_1+ \NLO_1\,, \nonumber \\
& {\rm NLO~QCD+EW} = \LO_1 + \NLO_1 + \NLO_2\,,\nonumber
\end{align}
where $\LO_2$ has been put in parentheses since it is numerically zero when the signature \eqref{signature} is considered.

It is worth to note that going beyond NLO for $pp \to \nu_e J J J$ production, in particular at $\ord(\alpha_s^3 \alpha^3)$, top-quark pair production with semi-leptonic decays is present and also contributes to the signature \eqref{signature}. Moreover,  it represents the largest contribution to the background in the searches for $t$-channel single-top production, see {\it e.g.} Ref.~\cite{Aaboud:2017pdi}.
This contribution appears only beyond the formal accuracy of our calculation and therefore it is not entering our results. However, it has to be taken into account for a correct estimate of the background.

\subsubsection{Input parameters}
\label{sec:input}

 In order to perform the calculation, given the presence of intermediate resonances, we use the complex-mass scheme.
 We use as input parameters for the EW sector $G_\mu$, $m_Z$ and $m_W$ and we accordingly perform the renormalisation in the $G_\mu$-scheme.
 The results of Sec.~\ref{sec:fixres} are obtained with the following masses
 and widths for the input parameters of the complex mass scheme\footnote{The same decay widths are used for LO and NLO calculations.}
\begin{align}
  &m_{Z}=91.1876~\textrm{GeV},&   &m_{W}=80.385~\textrm{GeV},&    &m_H=125~\textrm{GeV},&  &m_{\textrm{t}}=173.34~\textrm{GeV},&\\
   &\Gamma_{Z}=2.4955~\textrm{GeV},&    &\Gamma_{W}=2.0897~\textrm{GeV},&  &\Gamma_H=4.07~\textrm{MeV},&   &\Gamma_{\textrm{t}}=1.36918~\textrm{GeV}\, .&
\end{align}
In our calculation, the width of the Higgs boson is necessary only for regulating the integrable singularity of the $s$-channel Higgs boson that can be present in one-loop diagrams. 

The value of $G_\mu$ is set equal to 
\begin{equation}
G_\mu=1.16639\times 10^{-5}~\textrm{GeV}^{-2}\, ,
\end{equation}
and the CKM matrix is set equal to the $3\!\times\! 3$ identity matrix.  We renormalise QCD interactions in the $\MSbar$  scheme and, as already mentioned, we use the 5FS.\footnote{The parameter \texttt{SeparateFlavourConfigurations} has been introduced in {\sc MadGraph5\_aMC@NLO} in order to plot each one of the flavour configurations  independently, even if they are summed together because they have  identical matrix elements. This allows, for example, $b$-tagging  in the 5FS.} We set the renormalisation and factorisation scales to $\mu_R=\mu_F=H_T/2$, where $H_T$ is the scalar sum of the transverse momenta of  all the final-state particles, which are all massless. As PDF set we use {\sc\small LUXqed17\_plus\_PDF4LHC15\_nnlo\_100} \cite{Manohar:2016nzj, Manohar:2017eqh}, which
includes a photon member and $\alphas(m_Z)=0.118$. Scale uncertainties are evaluated via the standard 9-point independent variations of the factorisation and renormalisation scales.

\subsubsection{Clustering procedure}
\label{sec:clustering}
Since we perform a fixed-order NLO computation, at most two particles can be clustered together to generate signature \eqref{signature}.
Therefore, in order to ensure IR safety we perform the following procedure.
First of all we perform a QED clustering among leptons and photons, where $\Delta R=\sqrt{(\Delta \eta)^2 +(\Delta \phi)^2 }$ is the distance among two particles and the clustering parameter is set to $\Delta R^{\rm QED}=0.1$. In practice, we apply the anti-$k_T$ clustering algorithm with $p_{T, \rm min}^{\rm QED}=0$ GeV. If only a single particle is present within a radius  $\Delta R=\Delta R^{\rm QED}$, we do not cluster this particle with any other. Otherwise, in order to be IR safe, if the distance between two particles is $\Delta R<\Delta R^{\rm QED}$ we follow this procedure:
\begin{itemize}
\item If they are two photons,  they are not clustered.
\item If they are one photon and one lepton with flavour $f$, they are clustered and defined as a lepton with flavour $f$.
\item If they are an opposite-sign same-flavour (OSSF) lepton pair, they are clustered and defined as a photon.
\item If they are two leptons which are not  OSSF,  they are not clustered.
\end{itemize}

After this clustering procedure we can already reject events with three or more QED particles (leptons or photons). Indeed, leptons will not enter the QCD clustering and in order to obtain the signature \eqref{signature} we need at least a $\nu_e$, a $b$ or $\bar b$ and a positron. Therefore, four QED particles among four/five particles in the final state are not possible. Three would be in principle possible in the case of a positron and two photons, which in turn would lead to a light jet via the QCD clustering procedure discussed in the next paragraph. However,  the process $pp \TO e^+ \nu_e b \gamma \gamma $ is not possible due to flavour.

Then, we proceed with the standard QCD clustering for quarks, gluons and
photons in order to define QCD jets. Since we will perform the calculation for
the fiducial region defined in Ref.~\cite{Aaboud:2017pdi}, which is also
explicitly reported in Sec.~\ref{sec:signal},   we use the anti-$k_T$
clustering algorithm with parameters $\Delta R^{\rm QCD}=0.4$ and $p_{T,{\rm
    min}}^{\rm QCD}=30$ GeV. Only jets that contain a $b$ quark or antiquark
and have pseudorapidity $|\eta(j)| < 2.5$ are identified as $b$-jet with a
$100\%$ tagging efficiency. Note that in the case of $|\eta(j)| > 2.5$ or if a $b \bar b$ pair is present\footnote{In the analysis of Sec.~\ref{sec:shower}, where shower effects are taken into account, if a jet includes a $b \bar b$ pair it is considered as a $b$-jet.}, the jet is considered as a light jet. The latter condition is needed to ensure IR safety in the 5FS.

\subsection{Shower effects matched to NLO QCD}
\label{sec:showersetup}
In this section we describe the calculation of the contributions for Single-Top and $W$+jets productions to the signature \eqref{signature}, at NLO QCD accuracy and including QCD shower effects. Numerical results are presented later in Sec.~\ref{sec:shower}.

First of all, it is important to note that at the moment the theoretical knowledge for a consistent matching of NLO QCD and EW effects to (QCD) shower effects is not yet available for processes involving QCD interactions at LO.\footnote{In the case of processes involving a single perturbative order at LO and without QCD interactions, such as Drell-Yan or Higgs-Strahlung, this technology is available \cite{Barze:2012tt, Barze:2013fru, Granata:2017iod, Chiesa:2019ulk}. } The first attempts in this direction have been performed in refs.~\cite{Kallweit:2015dum, Gutschow:2018tuk, Granata:2017iod}. However, the approach pursued in these works applies only to the cases where EW effects are dominated by purely weak effects, in particular electroweak Sudakov logarithms. As we will explain in Sec.~\ref{sec:fixres}, this is not the case for the calculation performed in this work, both at the inclusive and differential level. In particular, in our calculation QED effects cannot be neglected due to the effective jet veto from the definition of the signature \eqref{signature}. Moreover, as we will see in Sec.~\ref{sec:shower}, QCD shower effects are large and so an analogous feature is expected, although at a smaller extent, also for EW corrections, which for these observables are mainly of QED origin. 

In this work therefore we consider only NLO QCD corrections to separately Single-Top and $W$+jets production and we match them to QCD parton shower effects. On the other hand, it is important to note that while such a calculation can be straightforwardly performed for the case of $W$+jets via public tools, in the case of Single-Top the situation is much more complex. Indeed, NLO QCD corrections to Single-Top correspond to the $\NLO_3$ contribution. The $\NLO_3$ involves both ``genuine'' QCD corrections on top of $\LO_3$, the LO in Single-Top, but also ``genuine'' EW corrections on top of $\LO_2$. Indeed the $\LO_2$ for the full process \eqref{syntax} is not vanishing, and therefore the IR structure of the $\NLO_3$ does involve QED singularities on top of the $\LO_2$. 

 On the other hand,  as also shown later in Sec.~\ref{sec:fixres}, the $\LO_2$ contribution is exactly zero when the signature \eqref{signature} is considered. Therefore, QED soft and collinear singularities are not involved in the $\NLO_3$ calculation for the signature \eqref{signature}; soft and collinear enhanced contributions in the $\NLO_3$ are only of QCD origin. For this reason, NLO QCD corrections can be matched to the QCD parton shower following the standard approaches \cite{Frixione:2002ik, Frixione:2007vw}.  On the other hand, since as we said  $\NLO_3$ in \eqref{syntax} involves QED singularities on top of the $\LO_2$, for the shower simulation we actually consider the process 
\begin{equation} \label{syntax_shower}
pp \to \nu_e b J J + pp \to \nu_e \bar b J J\, ,
\end{equation}
forcing one of the parton in the final state to be a bottom quark or antiquark. In this way, partonic channels that would be divergent due to QED interactions but that do not contribute to the signature \eqref{signature} are discarded from the beginning. We remind the reader that, in order to preserve IR safety, this approach cannot be pursued for the entire complete-NLO calculation described in Sec.~\ref{sec:theory} and results in  Sec.~\ref{sec:fixres}, and in general for other processes \cite{Frederix:2018nkq}. 

Results based on this approximation are presented in Sec.~\ref{sec:shower} and are based on the same input parameters, scale definitions and PDFs listed in Sec.~\ref{sec:input}. Only two differences are present. First,  photon-induced contributions have been ignored, but their contribution is negligible. Second, when a $b\bar b$ pair is clustered within a jet, this jet is tagged as a $b$-jet, at variance with the fixed-order calculation where it is instead tagged as a light jet.

 It is important to note that at LO (where the IR-safety problem cannot be present) in Single-Top numerical differences between the syntaxes  \eqref{syntax} and \eqref{syntax_shower} are completely negligible;  also when showering the events. Indeed, at  $\LO_3$ $J$ cannot be a gluon that the shower will subsequently split into a $b\bar b$ pair. Moreover, we have explicitly verified that also tagging a clustered $b\bar b$ pair as a light jet the NLO QCD predictions including shower effects are in general not affected above the percent level. Among all the plots presented in Sec.~\ref{sec:shower}, only for the $\mebj$ distribution above $150$ GeV differences of order $5\%$ are observed. 

In the case of $W$+ jets production, tagging a clustered $b\bar b$ pair as a
light jet has instead a non-negligible  impact. We have verified that this
reduces the NLO QCD predictions including shower effects by $\sim 10\%$ at the inclusive level. Nevertheless, in experimental analyses this procedure is typically not employed and a $b$-jet can contain more than a $b$-hadron.

\section{Results}
\label{sec:results}
\subsection{Fiducial region}
\label{sec:signal}

In order to isolate the contribution to the signature \eqref{signature} and select the fiducial region for $t$-channel single-top production we perform the following procedure at the analysis level, adopting the cuts from Ref.~\cite{Aaboud:2017pdi}.

As already mentioned in Sec.~\ref{sec:clustering} jets are clustered via the anti-$k_T$  algorithm with parameters $\Delta R^{\rm QCD}=0.4$ and $p_{T,{\rm min}}^{\rm QCD}=30$ GeV.  Also, only jets that contain a $b$ quark or antiquark and have pseudorapidity $|\eta(j)| < 2.5$ are identified as $b$-jets; in the case of $|\eta(j)| > 2.5$ a jet is always considered as a light jet. We also remind the reader that in the case of the fixed-order results,  if a $b \bar b$ pair is clustered, the corresponding jet is always considered as a light jet for IR safety. When we perform the calculation including shower effects this requirement is not necessary and therefore we consider such a jet a $b$-jet, still only if $|\eta(j)| < 2.5$. As already mentioned, we explicitly verified that this choice, being preferable because it is much closer to a realistic experimental procedure, has a negligible impact on the Single-Top results presented in this work. 

After having defined jets (and dressed leptons), we define the fiducial region according to \eqref{signature}, {\it i.e.}, by requiring exactly one light jet ($\ljet$), one $b$-jet ($\bjet$), a positron and missing transverse-energy. In particularly, following Ref.~\cite{Aaboud:2017pdi}, these cuts are applied:
\begin{itemize}
\item exactly one lepton: $|\eta(\ell)| < 2.5$ and  $p_T(\ell) > 25$ GeV and
  identified as a positron,
\item exactly one light jet: $|\eta(\ljet)| < 4.5$  and  $p_T(\ljet) > 30$ GeV,
\item exactly one $b$-jet: $|\eta(\bjet)| < 2.5$ and $p_T(\bjet) > 30$ GeV,
\item missing transverse-energy: $\met > 30$ GeV,
\item positron and jets separation: $\Delta R(e^+,\ell) > 0.4$,
\item positron and $b$-jet system: $\mebj<160$ GeV,
\end{itemize}
where  $\met \equiv p_T(\nu_e)$. 

The requirement of exactly two jets of which one being a light jet and one being a $b$-jet is suppressing the relative contribution of all the resonant processes besides the $t$-channel single top. Indeed, $s$-channel single top typically leads to two $b$-jets and $tW$ associate production to three jets. Also, $WZ$ and $W$ + jets production mostly lead to 2 $b$-jets or 2 light jets.

\subsection{Fixed order}
\label{sec:fixres}

In this section we present and discuss fixed-order results at complete-NLO accuracy for the total cross section and the differential distributions at 13 TeV;  we consider the signature \eqref{signature} in the fiducial region defined in Sec.~\ref{sec:signal}. As summarised in \eqref{conversion}, we will refer to the perturbative orders $\LO_3$, $\NLO_3$ and $\NLO_4$ as ``Single-Top'', while the remaining perturbative orders $\LO_1$, $\LO_2$, $\NLO_1$ and $\NLO_2$ will be referred as ``$W$+jets''.

One should bear in mind that although the fiducial region has been designed for enhancing the relative contribution of $t$-channel single-top, $s$-channel single-top and $tW$ contributions are still present and are not negligible. Numerical results on the relative impact of these other production channels and their dependence on the cuts are presented in Appendix \ref{sec:comparison}. Furthermore, considering directly the signature \eqref{signature}, the different processes cannot be separated in a gauge-invariant way at NLO and also non-resonant contributions are present, which also cannot be excluded for the same motivation. For this reason we have not subtracted the $s$-channel single-top and $tW$ contributions from the Single-Top predictions ({\it cf.}, the lower two diagrams in Fig.~\ref{fig:diagrams}).

We remind the reader that NLO (and also NNLO) QCD corrections to $t$-channel single-top with leptonic top decays have already been calculated  in a similar fiducial region in refs.~\cite{Berger:2016oht,Berger:2017zof}. However, this calculation is based on the narrow-width approximation for the top decay, therefore non-resonant effects, and $s$-channel and  $tW$ contributions are not taken into account. In Appendix \ref{sec:comparison} we perform a detailed comparison with the results in refs.~\cite{Berger:2016oht,Berger:2017zof}, showing how these effects can have an impact and motivating the features that are found in the results presented in this section, especially at the differential level.

\subsubsection{Total cross section}
\label{sec:total_fix}

In Tab.~\ref{table:FO} we present predictions for the signature \eqref{signature} in the fiducial region defined in Sec.~\ref{sec:signal}.
For different accuracies, we provide the results for the central value of the factorisation and renormalisation scale together with the associated scale uncertainties.\footnote{Scale uncertainties are evaluated via the standard 9-point independent variations of the factorisation and renormalisation scales.} We also show the relative size of QCD and EW corrections for the central values of the scales.
As can be seen, for Single-Top, the NLO QCD cross section ($\LO_3+\NLO_3$) is much smaller than the corresponding LO ($\LO_3$) prediction; the QCD $K$-factor is $\sim0.6$. This reduction is due to the requirement of exactly two jets; vetoing extra jets the QCD radiation is suppressed, yielding a negative correction. For the same reason, scale uncertainties do not strongly decrease moving from LO to NLO QCD accuracy and they are $\sim  8\%$ for the latter. However, in Sec.~\ref{sec:shower} we will see that taking into account shower effects, and therefore the multiple emissions of partons,  NLO QCD corrections do significantly decrease scale uncertainties ({\it cf.}~Tab.~\ref{table:FOvsPS_xsecs}). Also, we will show  that, unlike the fixed-order case, including shower effects LO and NLO QCD scale-uncertainty estimates are compatible. Moreover, we will compare NLO QCD predictions with or without  shower effects and we will show they are compatible, at the inclusive level. Thus, although scale uncertainties are larger at fixed order, NLO corrections are still sensible and reliable, with the exception of specific phase-space regions that we will specify in Sec.~\ref{sec:shower}.  

The impact of NLO EW corrections ($\NLO_4$) on the NLO QCD prediction is sizeable; it reduces the cross section by $-3\%$. Even though this is within the scale uncertainties, the latter are significantly reduced by shower effects. Therefore, for a correct comparison between theory predictions and experimental measurements EW effects have to be taken into account.

\begin{table}[t]
\begin{center}
\begin{tabular}{c  c}
\toprule
Single-Top & cross section \\
\midrule
LO & $4.623(1)_{ -0.533(-11.5 \%)}^{+ 0.415(+9.0 \%)}$ pb\\
NLO QCD & $2.762(6)_{ -0.240(-8.7 \%)}^{+ 0.226(+8.2 \%)}$ pb \\
NLO QCD+EW & $2.676(6)_{ -0.236(-8.8 \%)}^{+ 0.229(+8.6 \%)}$ pb\\
(NLO QCD)/LO & 0.60(1)\\
(NLO QCD+EW)/(NLO QCD) & 0.97(1)\\
\bottomrule
\multicolumn{1}{c}{} &\multicolumn{1}{c}{} \\
\toprule
$W$+jets & cross section \\
\midrule
LO & $0.7656(6)_{ -0.2265(-29.6 \%)}^{+ 0.3002(+39.2 \%)}$ pb\\
NLO QCD & $1.612(3)_{ -0.309(-19.2 \%)}^{+ 0.323(+20.1 \%)}$ pb \\
NLO QCD+EW & $1.597(3)_{ -0.305(-19.1 \%)}^{+ 0.318(+19.9 \%)}$ pb\\
(NLO QCD)/LO & 2.11(1) \\
(NLO QCD+EW)/(NLO QCD) & 0.99(1) \\
\bottomrule
\end{tabular}
\end{center}
\caption{Various fixed-order cross sections (in pb), including their
  scale uncertainty, for the signature \eqref{signature} within the
  fiducial region defined in Sec.~\ref{sec:signal} for the Single-Top
  process (top table) and the $W$+jets process (bottom table). The
  ratios (last two lines of both tables) are computed for the central
  values of the corresponding predictions.}
\label{table:FO} 
\end{table}

In Tab.~\ref{table:FO}, we also show results for $W$+jets, {\it i.e.}, the contributions from the remaining perturbative orders $\LO_1$, $\LO_2$, $\NLO_1$ and $\NLO_2$.
The NLO QCD cross section ($\LO_1+\NLO_1$) is much larger than the
corresponding LO ($\LO_1$) prediction; the QCD $K$-factor is $\sim2.1$. Unlike
the case of Single-Top, the requirement of exactly two jets does not lead to
negative corrections. This pattern is unusual for a NLO QCD calculation with a
requirement of an exclusive number of jets, {\it i.e.}, applying a
jet-veto. However, in this process real QCD radiation can convert LO events
that would not contribute to the signature \eqref{signature} in events that do
contribute.  For example, $e^+\nu_e g g$ final states, which are present at
LO, do not contribute to the signature \eqref{signature}. On the other hand,
real QCD radiation can convert them via the $g\TO b\bar b$ splitting into a
$e^+\nu_e g b \bar b$ final state, which can contribute to the signature
\eqref{signature}. Moreover, the LO $e^+\nu_e g g$ final state has a much
larger cross section than the $e^+\nu_e b q$ one, which does contribute to the
signature \eqref{signature} at LO. Hence, the NLO QCD contributions increase
the central value of the LO cross section by more than a factor 2.

At variance with Single-Top predictions,  scale uncertainties  decrease moving from LO ($\sim{}^{+40\%}_{-30\%}$) to NLO QCD ($\sim{}^{+20\%}_{-20\%}$) accuracy. However, despite this reduction,  they are  larger than in the case of Single-Top, due to the higher powers of $\alpha_s$ factorising the LO prediction.  On the other hand, similarly to the case of Single-Top, LO and NLO QCD predictions do not overlap. We will show in Sec.~\ref{sec:shower} that including shower effects NLO QCD scale uncertainties are strongly reduced and are compatible with the fixed-order case.

The impact of NLO EW corrections ($\NLO_2$) on the NLO QCD prediction is
instead negligible at the inclusive level; it reduces the cross section by
$-1.0\%$, {\it i.e.},  NLO EW corrections are much smaller than scale
uncertainties. The $\LO_2$ is instead exactly equal to zero. Indeed, although the process $\eqref{syntax}$, { \it i.e.}, with fully democratic jets, involves non-vanishing contributions to the $\LO_2$, by requiring a $b$-jet in the analysis they all vanish.

\subsubsection{Differential distributions}
\label{sec:diff_fix}

We move now to differential distributions. Similarly to the inclusive case, we separate results for Single-Top and $W$+jets predictions, but one should bear in mind that both refer to the same final state. We provide at the end also a direct comparison at differential level for Single-Top and $W$+jets predictions.

\begin{figure}[!t]
\centering
\includegraphics[width=0.38\textwidth]{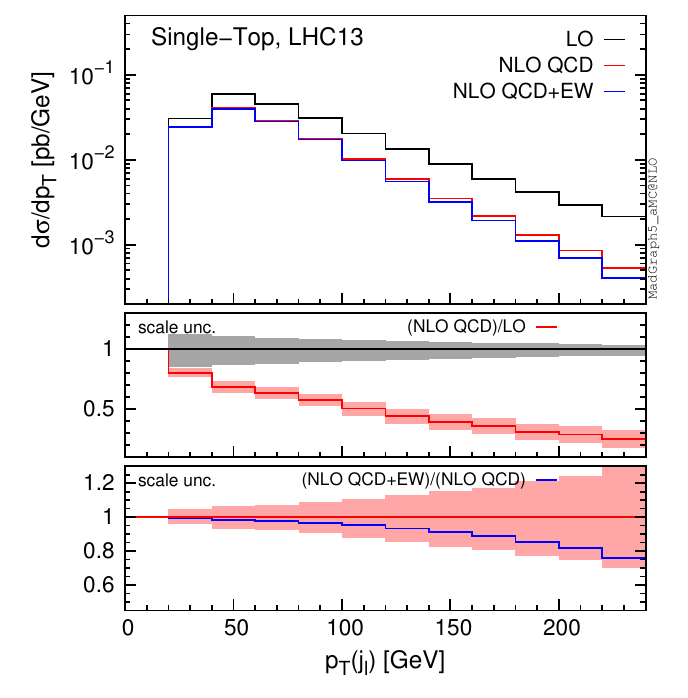}
\includegraphics[width=0.38\textwidth]{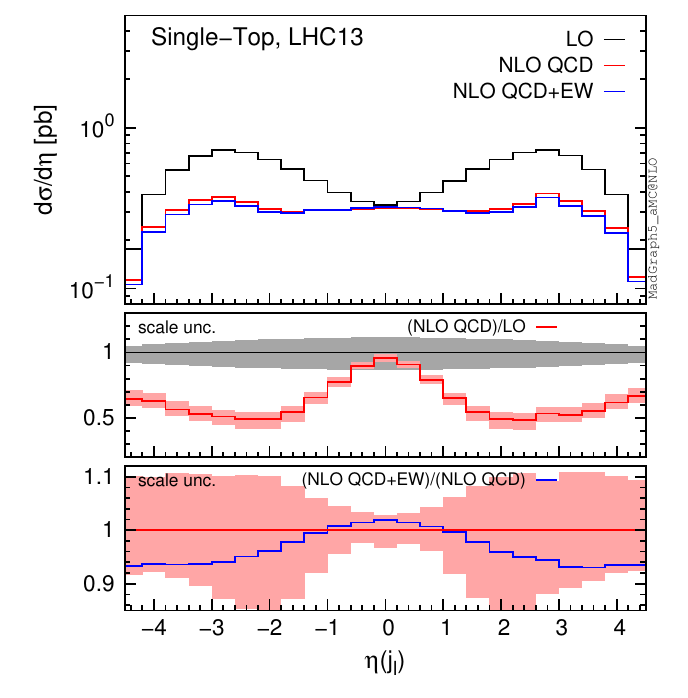}
\includegraphics[width=0.38\textwidth]{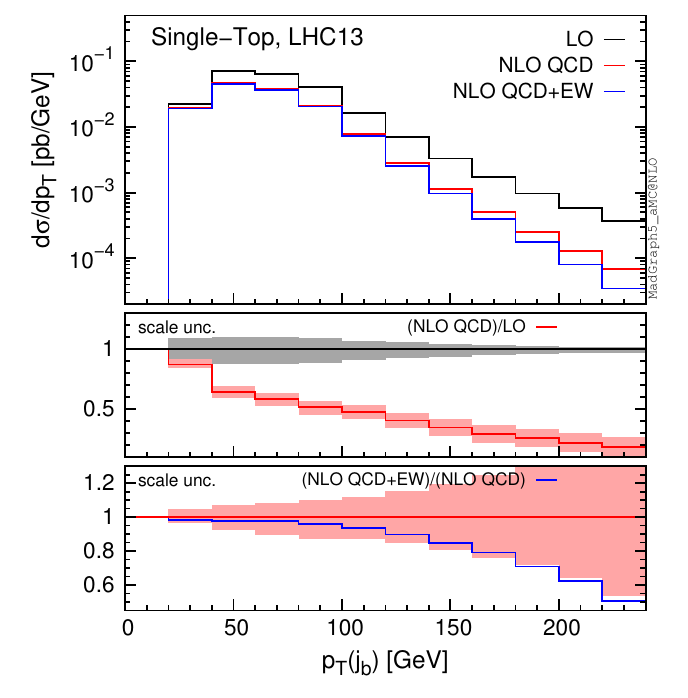}
\includegraphics[width=0.38\textwidth]{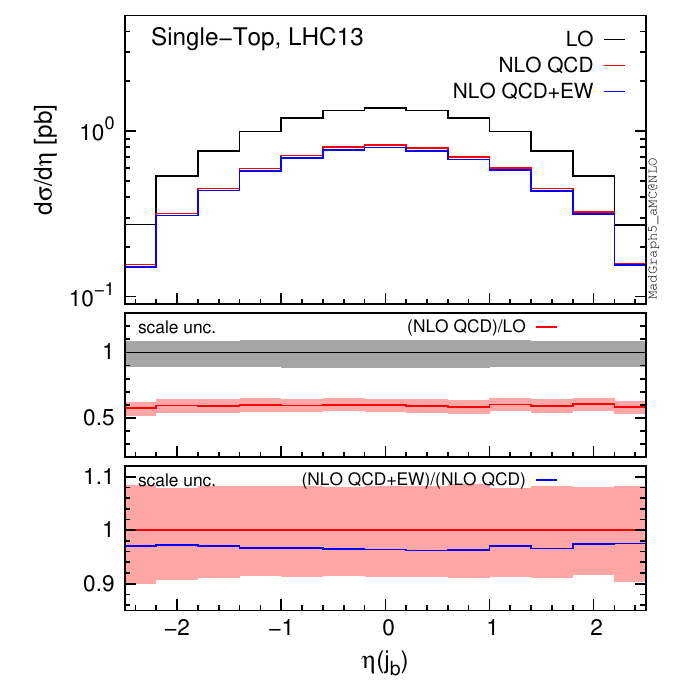}
\includegraphics[width=0.38\textwidth]{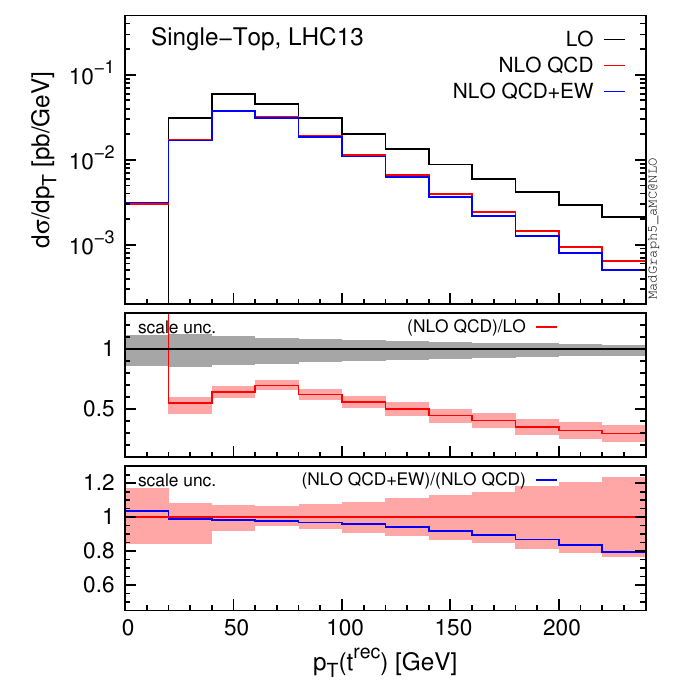}
\includegraphics[width=0.38\textwidth]{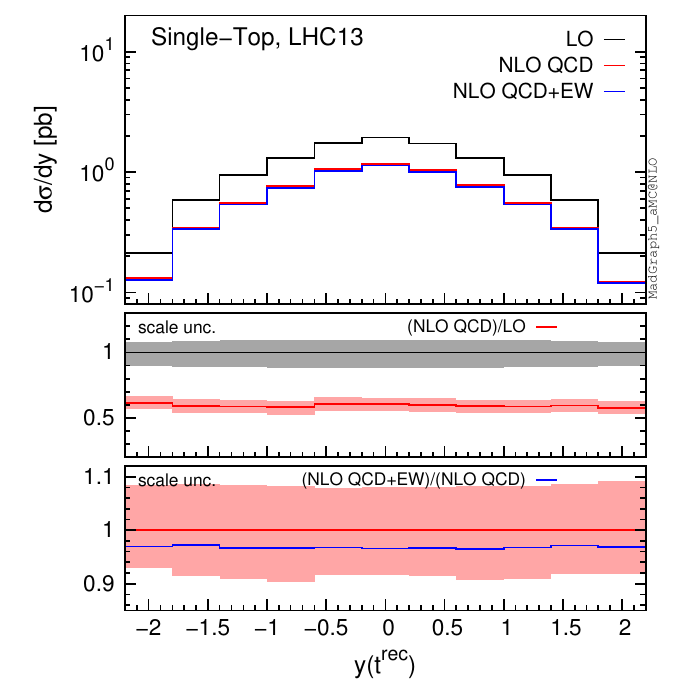}
\caption{LO, NLO QCD and NLO QCD+EW predictions for the transverse momentum (left) and rapidity (right) distributions of the light jet (top), $b$-jet (central) and reconstructed-top (bottom) for the Single-Top process. }
\label{fig:FO1}
\end{figure}

\begin{figure}[!t]
\centering
\includegraphics[width=0.38\textwidth]{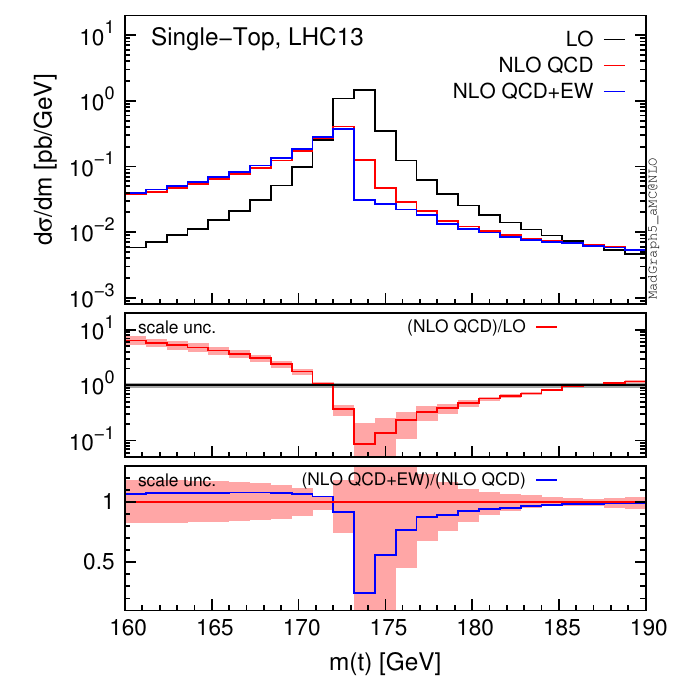}
\includegraphics[width=0.38\textwidth]{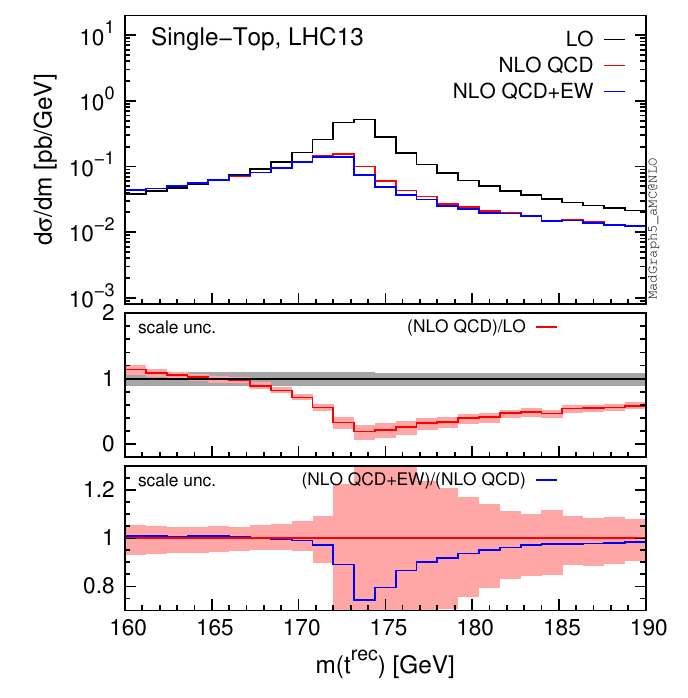}
\includegraphics[width=0.38\textwidth]{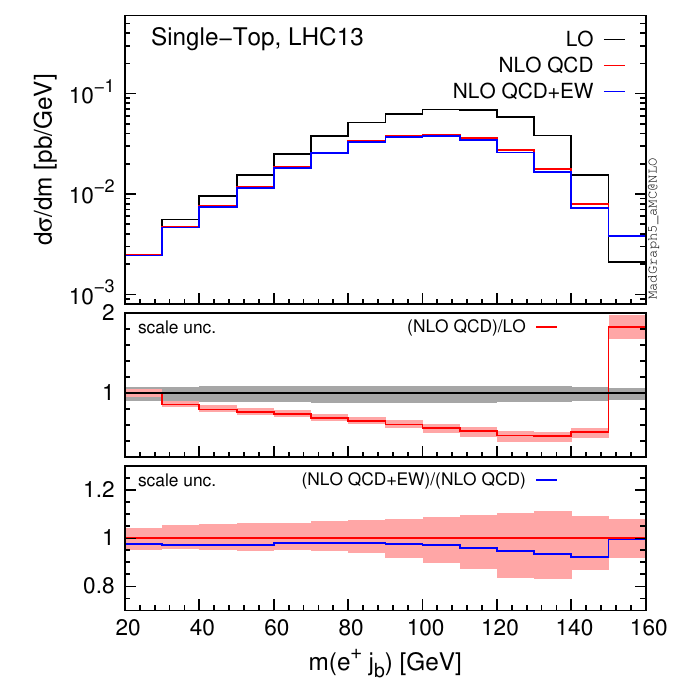}
\includegraphics[width=0.38\textwidth]{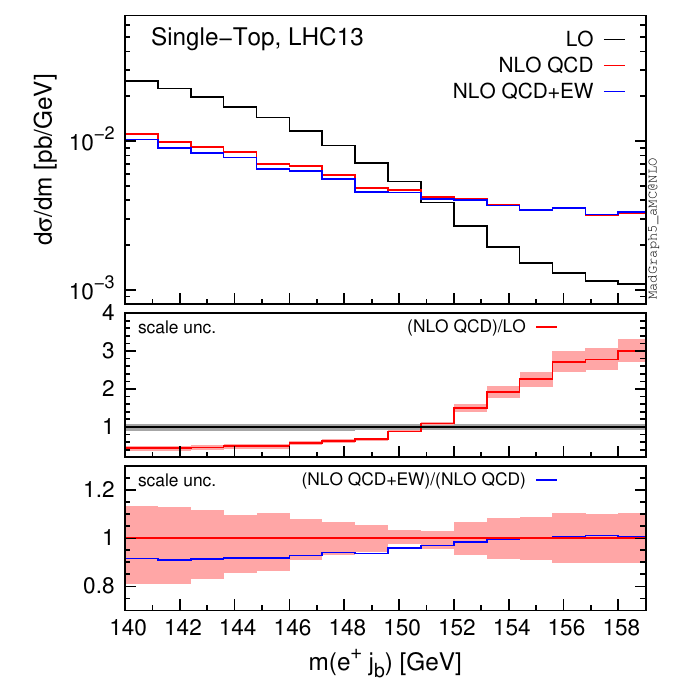}
\includegraphics[width=0.38\textwidth]{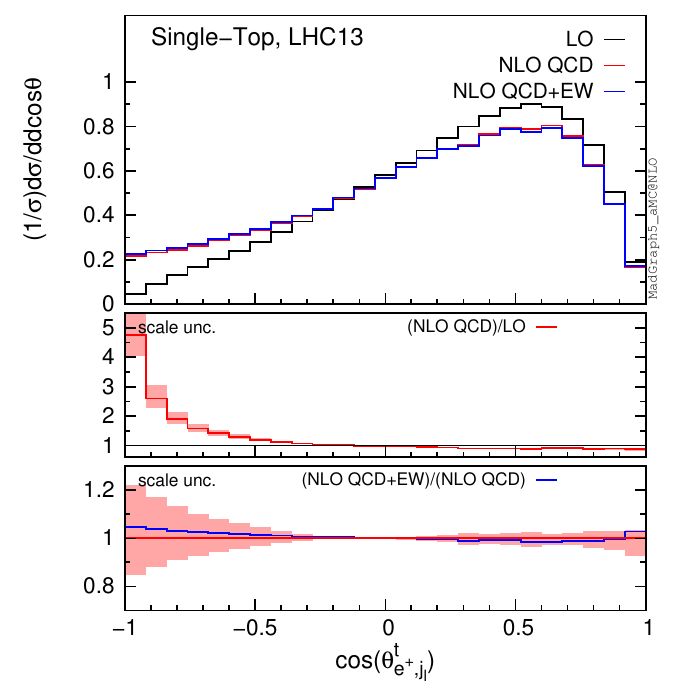}
\includegraphics[width=0.38\textwidth]{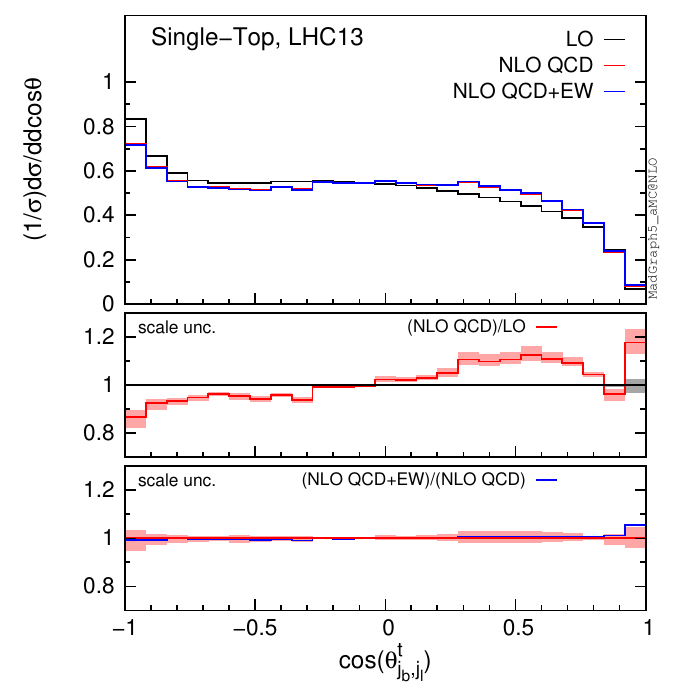}
\caption{LO, NLO QCD and NLO QCD+EW predictions for the top-quark invariant mass (top) at the truth level (left) and reconstructed (right), positron and $b$-jet system invariant mass distribution (central) in a large (left) and small range (right), and spin-correlation observables (bottom) for the Single-Top process.}
\label{fig:FO2}
\end{figure}

 In Figs.~\ref{fig:FO1} and \ref{fig:FO2} we show predictions for several quantities, taking into account the cuts of Sec.~\ref{sec:signal} that define the fiducial region. All plots display results for Single-Top contributions at LO, NLO QCD and NLO QCD+EW accuracy. In the first inset we show both LO (black) and NLO QCD (red) predictions including scale uncertainties normalised to the central value of the LO; the latter is the QCD $K$-factor. In the second inset we show again the NLO QCD scale uncertainties, but now normalised to the central NLO value, and also the  (NLO QCD+EW)/(NLO QCD) ratio.
In Fig.~\ref{fig:FO1} we display the predictions for  the transverse momentum of the light jet ($\ljet$), the $b$-jet ($\bjet$) and the reconstructed momentum of the top-quark $\ptrec$. We show also the pseudorapidity for  the light jet and the $b$-jet, and the rapidity for the reconstructed top. In Fig.~\ref{fig:FO2} we instead show the predictions for the quantities $\mttrue$, $\mtrec$, $\mebj$, $\coselj$ and $\cosbjlj$, which we define in the following. The quantity $\mttrue$ is the invariant mass of the positron, the $b$-jet and the momentum of the neutrino and therefore, although it cannot be directly measured, corresponds to the true momentum of the would-be top-quark. On the contrary, $\mtrec$ is the same quantity,  but with the momentum of the neutrino  extracted from the value of $\met$ assuming the $W$ boson being on-shell. In practice, one has to solve the quadratic equation $ m_W^2= 2 p_{e^+} \cdot p_{\nu_e}$ for $p^z_{\nu_e}$ assuming $\met=p_T(\nu_e)$.\footnote{We use the same procedure employed in Ref.~\cite{Sirunyan:2016cdg}. We select the solution that is the smallest in absolute value. If the two solutions are complex, we rescale the $p_T(\nu_e)$ components such that we get one real  solution.} The same procedure is used also for the determination of $\ptrec$. The quantity $\mebj$ is the invariant mass of the positron and $b$-jet, which is exploited for the measurement of $m_t$. Indeed, at LO and assuming the top-quark and the $W$ boson on-shell, $\mebj$ has an end-point for $\mebj=\sqrt{m_t^2-m_W^2}\sim 154 ~ {\rm GeV}$. For this reason in Fig.~\ref{fig:FO2} we show this distribution both in a wide range (central-left plot) and close to the aforementioned end-point (central-right plot). The quantities $\coselj$ and $\cosbjlj$ are the cosine of the angle between the positron and the light jet and of the angle between the $b$-jet and light jet in the top-quark rest-frame, respectively~\cite{Frixione:2007zp}. Via these angular distributions it is possible to gain information on the top-quark polarisation along the direction of the spectator light jet $(\ljet)$ that is present in the $t$-channel production. The positron angular distribution $\coselj$ carries the higher spin-analysing power (degree of correlation)  with the top-quark spin. Thanks to the dependence on the top-quark spin and polarisation  these distributions are sensitive to new physics~\cite{deBeurs:2018pvs,Arhrib:2016vts,Jueid:2018wnj,Arhrib:2018bxc,Arhrib:2019tkr}.


Consistently with the result at the inclusive level, differential QCD $K$-factors of plots in Fig.~\ref{fig:FO1} are in general substantially smaller than one.
Especially, the $p_T(\ljet)$ and $p_T(\bjet)$ distributions exhibit very large and negative corrections in the tail; they are $\sim - 70 \%$ at 200 GeV and well outside the LO scale uncertainty. This feature is induced by the requirement of exactly two jets, and clearly shows the necessity of resumming large QCD Sudakov logarithms in this phase-space region. As we will see in Sec.~\ref{sec:shower} this effect is counterbalanced by QCD shower effects ({\it cf.}~upper-left plot in Fig.~\ref{fig:FO_PS_1}). The QCD $K$-factor is instead very different for $\eta(\ljet)$ and $\eta(\bjet)$; it is very flat for $\eta(\bjet)$ while $\eta(\ljet)$ shows a bump centred around $\eta(\ljet)=0$. As discussed in Appendix \ref{sec:comparison}, this effect is due to the large contribution from $tW^-$ production entering via NLO QCD corrections. Moreover, $\bar{t}W^+$ is also contributing, increasing even more this effect. Indeed, in $\bar t W^+$ production the $W^+$ boson can decay into $e^+\nu_e$, while the top can decay hadronically, contributing in total to the signature \eqref{signature} (see {\it e.g.}~bottom-right diagram of Fig.~\ref{fig:diagrams}). Considering EW corrections, for both $\eta$ and $p_T$  distributions of the light and $b$-jets,  the shape of the  (NLO QCD+EW)/(NLO QCD) ratio is very similar to the one of the corresponding QCD  $K$-factor. In particular, for $p_T(\ljet)~ (p_T(\bjet))\sim$  200 GeV, EW corrections further reduce the NLO QCD predictions by $\sim 20(30) \% $. Also, EW corrections similarly to QCD corrections do not lead to large effects for central light jets. The $p_T$ and $\eta$ distributions for the reconstructed top-quark have features very similar to the case of the $b$-jet, besides the region $\ptrec \leq $ 80 GeV. In all the aforementioned cases the NLO EW corrections are within the NLO QCD scale uncertainties. On the other hand, as already mentioned, shower effects significantly decrease scale uncertainties as discussed in Sec.~\ref{sec:diffshower} ({\it cf.} Fig.~\ref{fig:FO_PS_1}). 

We move now to the discussion of plots in Fig.~\ref{fig:FO2}. The $\mttrue$ distribution receives enormous corrections from QCD manifesting also as very large scale uncertainties that nevertheless do not overlap with the LO ones. This effect is induced by real radiation from the bottom quark which is not clustered into the $b$-jet and therefore leads to the migration of events from the LO peak $\mttrue\simeq m_t$ to smaller values. Moreover, this effect is further enhanced by the requirement of exactly   one $b$-jet and one light jet. Indeed, at NLO QCD, often these two jets corresponds to the $b$ quark and the unclustered gluon emitted by it, with instead the light-jet from $t$-channel production ($t\ljet$) being actually forward and undetected. Thus, the shape of the NLO QCD prediction and the QCD $K$-factor of $\mttrue$ strongly depend on the veto and the clustering radius $\Delta R^{\rm QCD}$. As a further check, we have investigated the effect of a jet veto on $\mttrue$ for $pp\TO W^+ b W^- \bar b$ production at NLO QCD accuracy. While without jet veto we find results qualitatively in very good agreement with the case of $\mttrue$ for a top quark with leptonic decays discussed in Ref.~\cite{Denner:2017kzu}, a veto on additional QCD radiation strongly affects the distribution and moves the position of the peak as in Fig.~\ref{fig:FO2}. The very large $K$-factor at $\mttrue<m_t$ is induced by the migration of events from the peak to the off-shell region.  

The situation is a bit different for the case of $\mtrec$. NLO QCD scale uncertainties are not as large as in the $\mttrue$ case and, especially, we do not see the very large $K$-factor at $\mtrec<m_t$, which we observe at $\mttrue<m_t$. Regardless of the value of $\mttrue$, most of the events are associated to an on-shell $W$, but nonetheless for a small fraction of them the reconstructed value of the longitudinal component of the neutrino momentum is different to the true value. This fraction is anyway sufficiently large to lead to a much flatter LO distribution for $\mtrec$ w.r.t.~$\mttrue$; this effect is due to events with $\mttrue \simeq m_t$ and $|\mtrec-m_t|\gg\Gamma_t$, which are a small fraction w.r.t.~the generic events with $\mttrue \simeq m_t$, but not w.r.t.~those with $\mttrue\simeq\mtrec$. When NLO QCD corrections are calculated, the migration effects from the peak of $\mtrec$ is then much smaller, and therefore we do not observe large $K$-factors for  $\mtrec<m_t$. Although smaller in size, a similar effect is observed also for EW corrections, which for both $\mttrue$ and $\mtrec$ are within the NLO QCD scale uncertainties.

The $\mebj$ distribution (central-left plot) receives also very large negative QCD corrections due to the jet veto, besides the last bin where only events originating from a off-shell top-quark and/or $W$ boson can contribute at LO. We zoom the last two bins in the central-right plot, where this effect can be seen even better in proximity to the value  $\mebj=\sqrt{m_t^2-m_W^2}\sim 154 ~ {\rm GeV}$. Electroweak corrections start being rather flat at $\sim -2\%$ in the $\mebj<100 ~ {\rm GeV}$ range. They increase in absolute value reaching a relative size of $\sim -10\%$ for  $\mebj \sim140 ~ {\rm GeV}$ and then quickly decrease up to the cut $\mebj=160 ~ {\rm GeV}$. Also for this observable parton-shower effects are expected to be non-negligible.
Finally, we discuss the $\coselj$ and $\cosbjlj$ normalised distributions, for which scale uncertainties are calculated via the envelope of the 9-point variation correlated in the numerator and the denominator. For values $\coselj<-0.5$, QCD corrections are positive and large (reaching a factor of $\sim 5$) and they are further increased by EW corrections, which are instead negligible over the rest of the phase space. On the contrary, in the case of the $\cosbjlj$ normalised distribution, EW corrections are negligible and QCD corrections are at most $\sim 15\%$ in absolute value. 

Summarising,  the plots in Figs.~\ref{fig:FO1} and \ref{fig:FO2} clearly show that EW effects are sizeable, although within the NLO QCD scale uncertainties, and that in very specific phase-space regions (the tails of the $p_T(\ljet)$ and $p_T(\bjet)$ distributions, the peak of the $\mttrue$ and $\mtrec$ and the $\mebj$ bin around the $\sqrt{m_t^2-m_W^2}\sim 154 ~ {\rm GeV}$ region) QCD effects are not under control at fixed order; NLO QCD and LO scale-uncertainty bands do not in general overlap.  Precisely for this reason, in Sec.~\ref{sec:shower} we analyse the impact of shower effects, which as anticipated reduce the impact of NLO QCD corrections and also the associated scale uncertainties. 

\medskip

\begin{figure}[!t]
\centering
\includegraphics[width=0.40\textwidth]{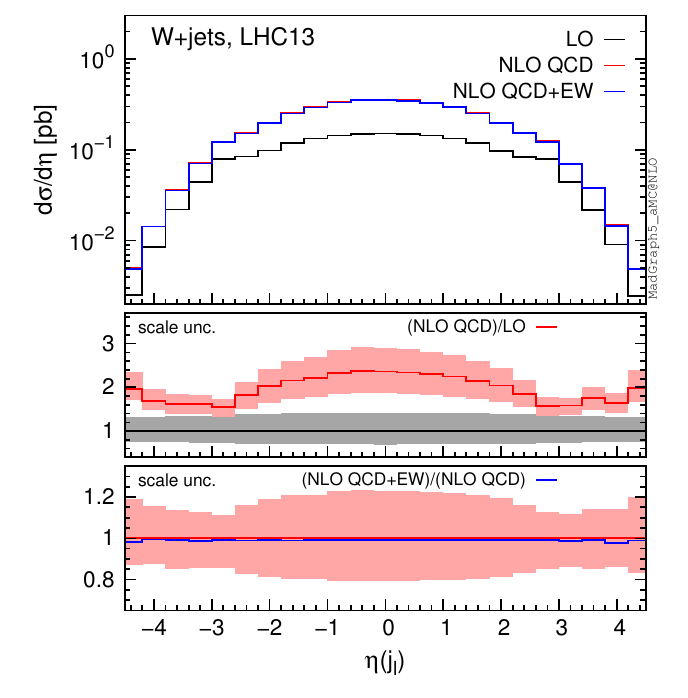}
\includegraphics[width=0.40\textwidth]{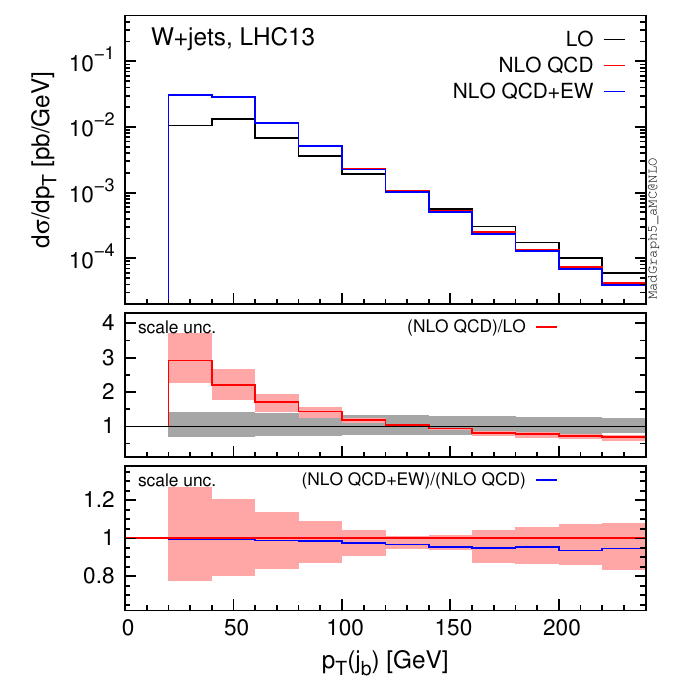}
\includegraphics[width=0.40\textwidth]{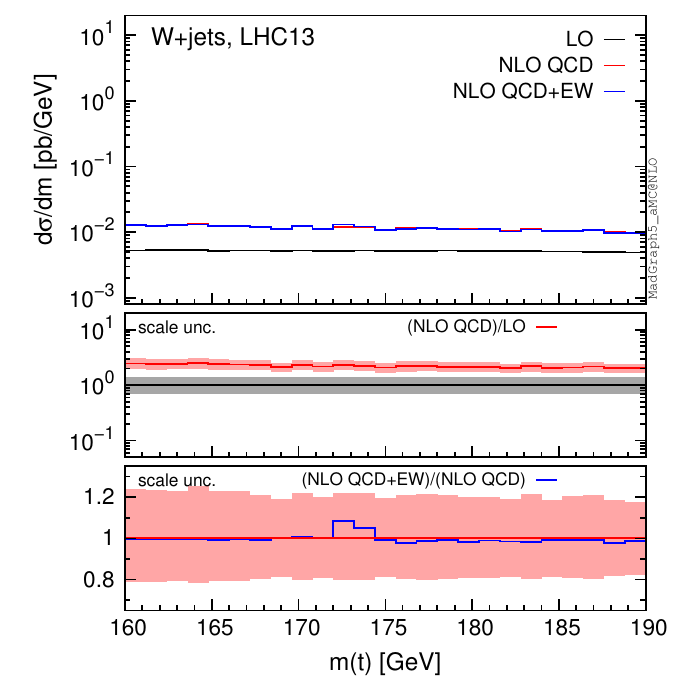}
\includegraphics[width=0.40\textwidth]{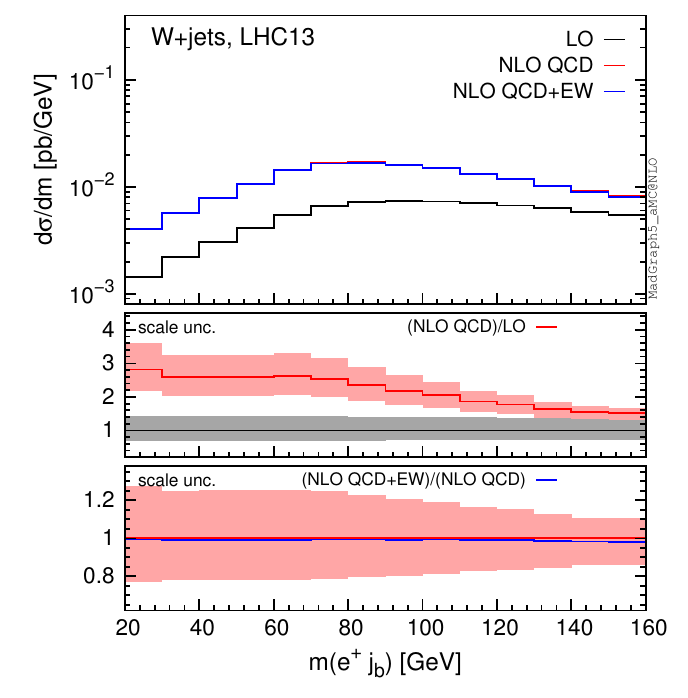}
\includegraphics[width=0.40\textwidth]{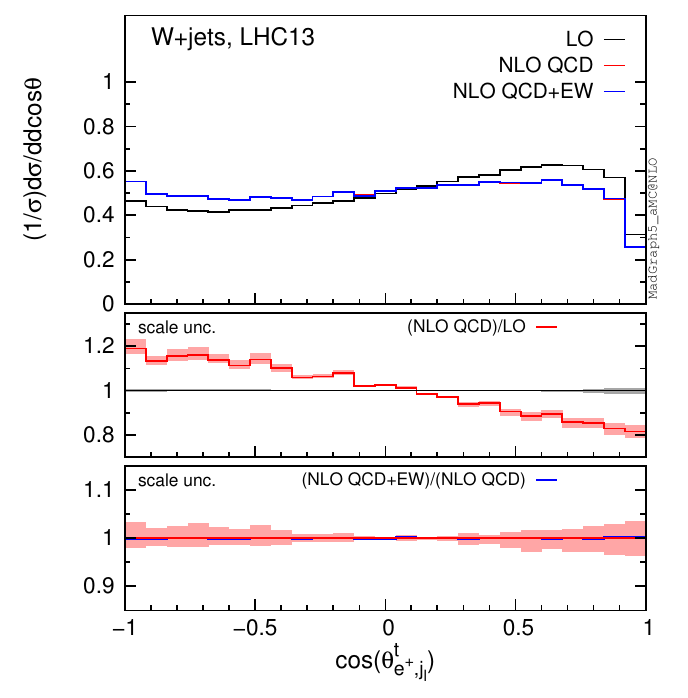}
\includegraphics[width=0.40\textwidth]{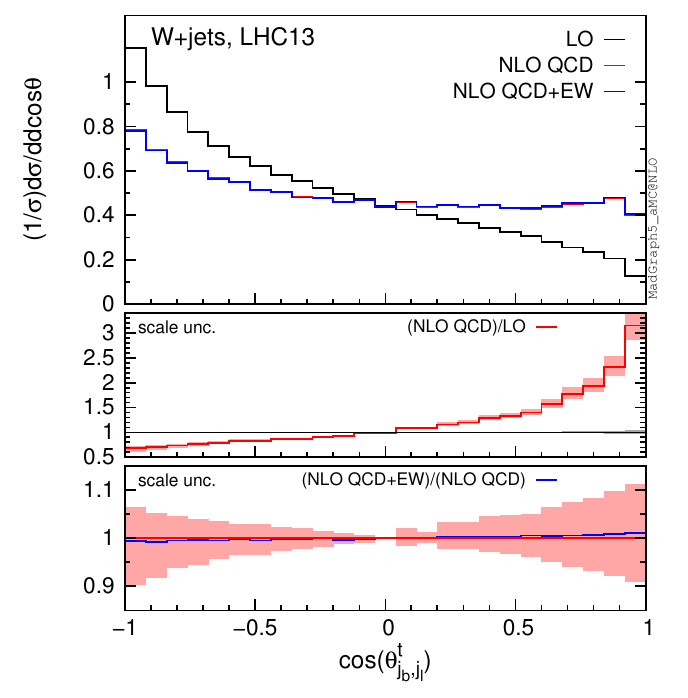}

\caption{LO, NLO QCD and NLO QCD+EW predictions for the observables from Figs.~\ref{fig:FO1} and \ref{fig:FO2} that show non-flat effects from either NLO QCD or NLO EW corrections for the $W$+jets process.}
\label{fig:FO3}
\end{figure}

In Fig.~\ref{fig:FO3} we show predictions at LO, NLO QCD and NLO QCD+EW accuracy for $W$+jets.  The layout of the plots is the same of those in Figs.~\ref{fig:FO1} and \ref{fig:FO2} and we have shown only the distributions that have already been considered for the case of Single-Top and that show non-flat effects from either NLO QCD or NLO EW corrections, {\it i.e.}, $\eta(\ljet)$,  $p_T(\bjet)$, $m(t)$, $\mebj$ in the large range, $\coselj$ and $\cosbjlj$. For all these observables, also in the $W$+jets case, NLO QCD scale uncertainties are large and are not compatible with the LO ones.

At variance with Single-Top, the light-jet and the $b$-jet distributions have a very different shape in the case of $W$+jets contribution to the signature \eqref{signature}.
Therefore, the fact that all jets with $|\eta|>2.5$ are tagged as light jets
has a strong impact  on the $\eta(\ljet)$ distribution (top-left
plot). Indeed, especially in the (NLO QCD)/LO ratio, this distribution shows a
very different behaviour for $|\eta(\ljet)|<2.5$ and $|\eta(\ljet)|>2.5$. The
NLO EW corrections are instead rather flat. Moving to $p_T(\bjet)$ (top-right
plot), the impact of NLO QCD corrections on  this distribution is very
different from the Single-Top case. At small $p_T(\bjet)$ values the QCD
$K$-factor is almost equal to 3 and becomes much smaller than 1 at large
$p_T(\bjet)$. Therefore, also in the case of $W$+ jets shower effects are
expected to be relevant. Results at the inclusive level including QCD shower effects are presented in Sec.~\ref{sec:totalshower} for $W$+jets. NLO EW corrections exhibit a negative growth for large $p_T$ values due to the jet veto, but for this observable are smaller in absolute value than in the case of Single-Top, reaching at most $-5\%$ in the tail. The $m(t)$ distribution and the corresponding NLO QCD corrections (center-left plot) are very flat in the range considered, since no top-resonance is present. However, the (NLO QCD+EW)/(NLO QCD) ratio is non-flat around  $m(t)\sim m_t$, with the typical shape induced by a resonance and has a $+10\%$ impact for $m(t)\lesssim m_t$. This effects is induced by interferences among real emission diagrams of  $\ord (\alphas^{3/2}\alpha)$ and $\ord(\alphas^{1/2}\alpha^{3/2})$, where the latter order contains diagrams with $s$-channel top-quark propagators and therefore induces this effect. On the other hand, in the case of $\mtrec$ we have checked that the non-flat effect is reduced a lot, being at most $2\%$.

For all the three remaining distributions in Fig.~\ref{fig:FO3}, NLO EW corrections are  small and flat w.r.t.~the NLO QCD predictions, while the QCD $K$-factor is not flat.
Since top-quark resonances are not present at LO or NLO QCD, the shape of the $\mebj$ distribution is completely different from the Single-Top case and consequently also the QCD $K$-factor, which is positive over the full $\mebj$ range considered and ranges form $\sim 3$ at  $\mebj=20$ GeV to $\sim 1.5$ at  $\mebj=160$ GeV.
The same argument applies also to the $\coselj$ and $\cosbjlj$ distribution. On the other hand, while in the case of $\coselj$ the QCD $K$-factor monotonically decreases from 1.2 at $\coselj=-1$ to 0.8 at $\coselj=1$, in the case of $\cosbjlj$ it monotonically grows from 0.7 at $\cosbjlj=-1$ to 3 at $\cosbjlj=1$ using our binning.

\medskip

\begin{figure}[!t]
\centering
\includegraphics[width=0.43\textwidth]{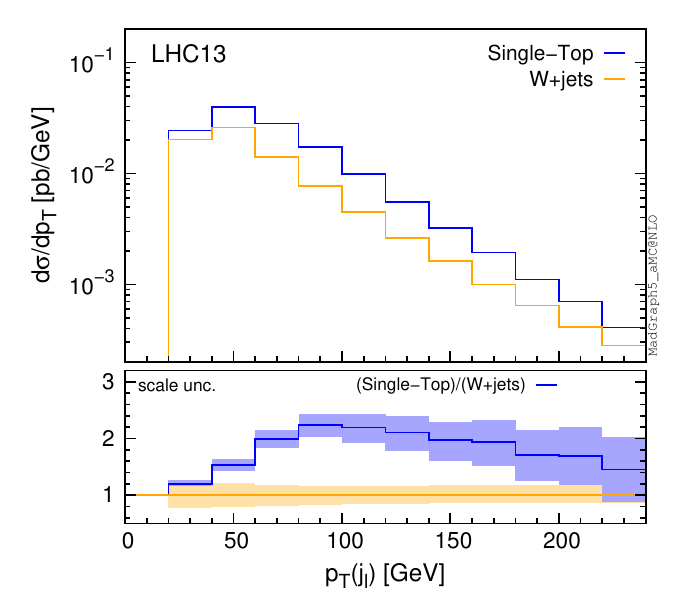}
\includegraphics[width=0.43\textwidth]{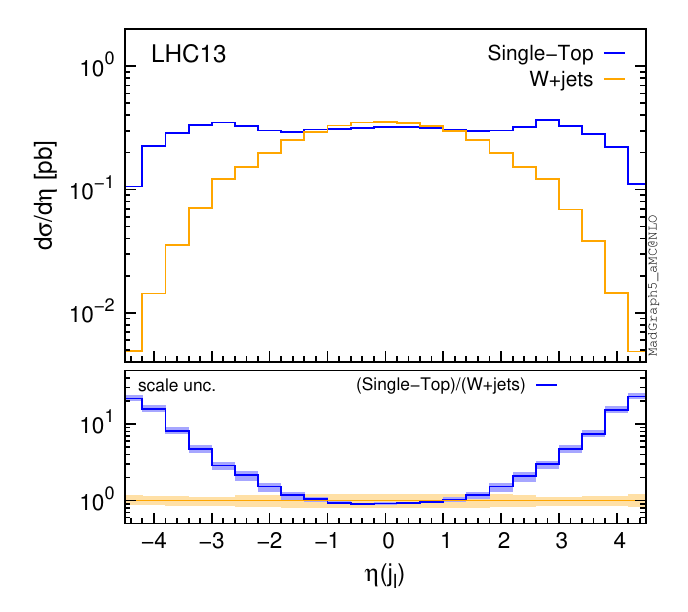}
\includegraphics[width=0.43\textwidth]{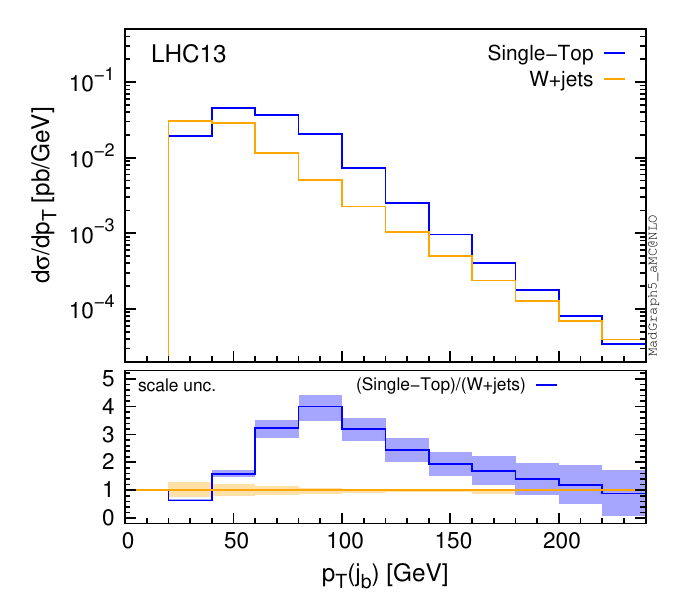}
\includegraphics[width=0.43\textwidth]{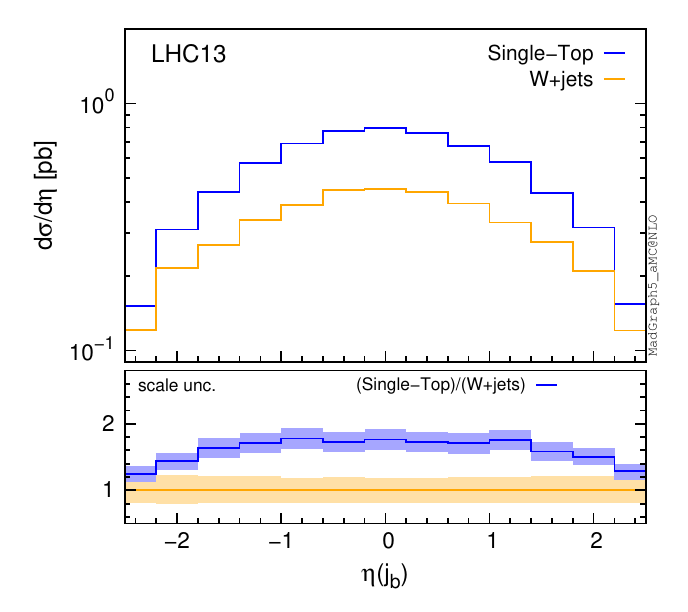}
\includegraphics[width=0.43\textwidth]{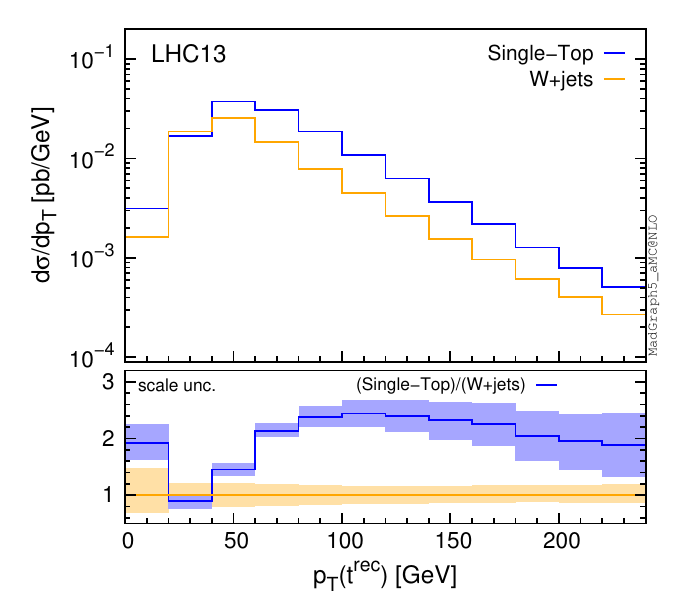}
\includegraphics[width=0.43\textwidth]{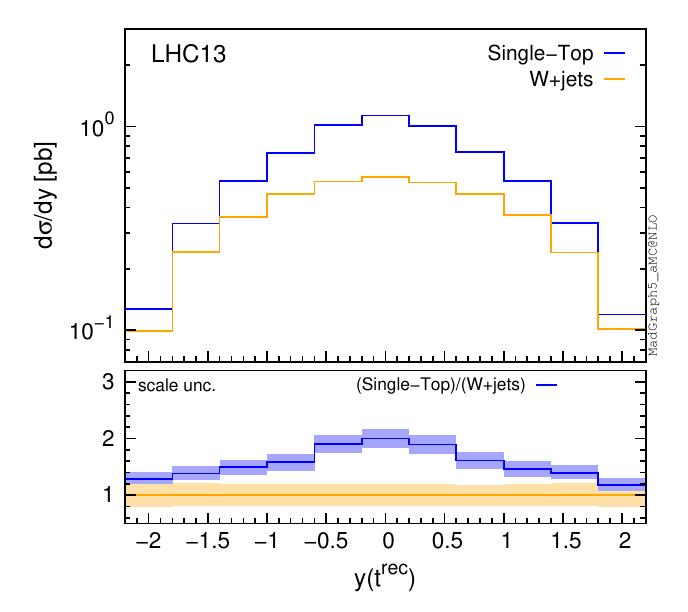}
\caption{Comparison between Single-Top and $W$+jets predictions at NLO QCD+EW accuracy for the same observables considered in Fig.~\ref{fig:FO1}.}
\label{fig:SB1}
\end{figure}

\begin{figure}[!t]
\centering
\includegraphics[width=0.43\textwidth]{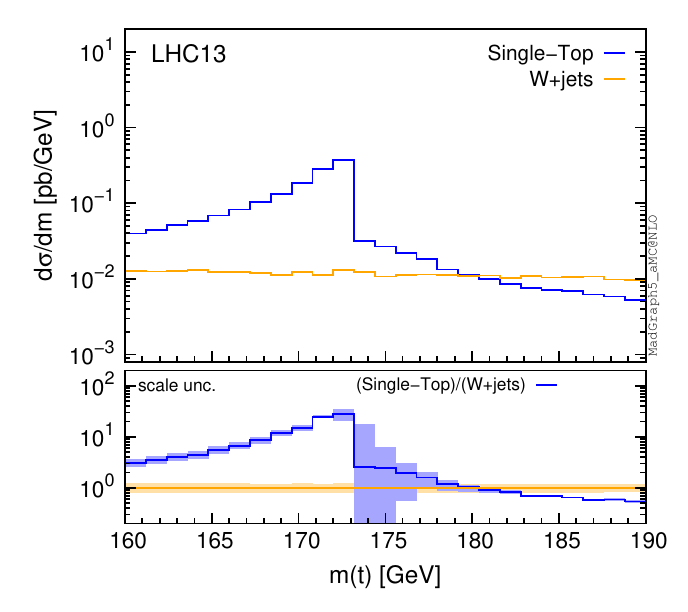}
\includegraphics[width=0.43\textwidth]{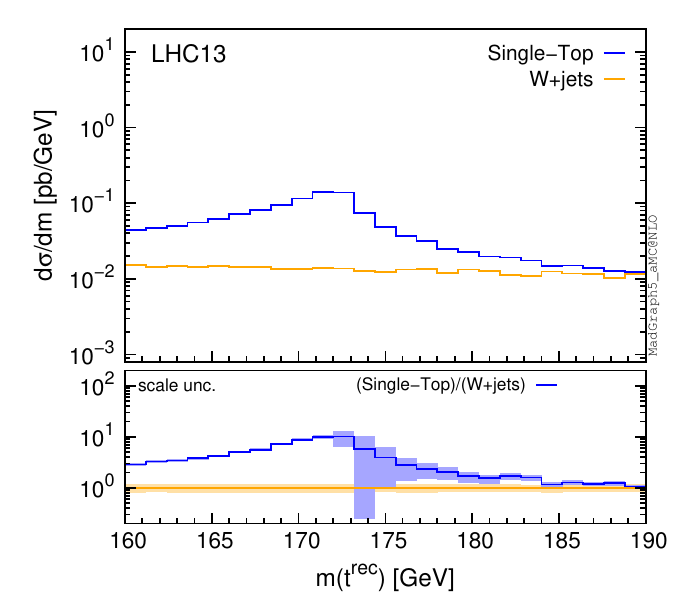}
\includegraphics[width=0.43\textwidth]{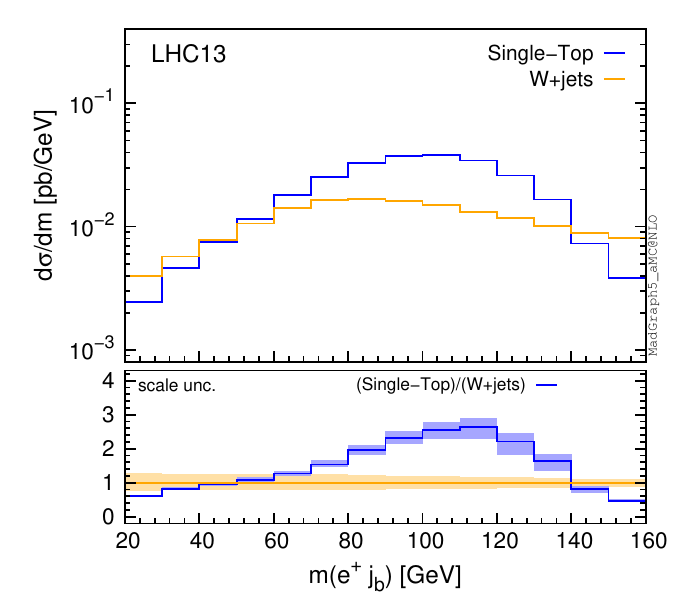}
\includegraphics[width=0.43\textwidth]{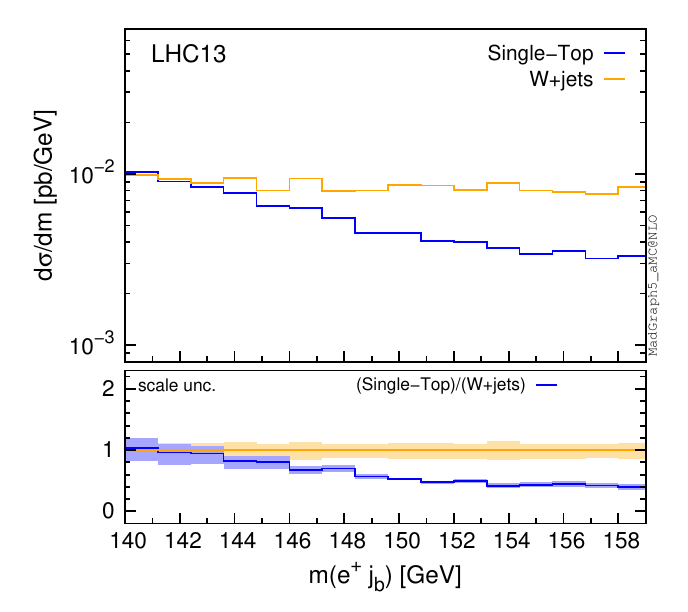}
\includegraphics[width=0.43\textwidth]{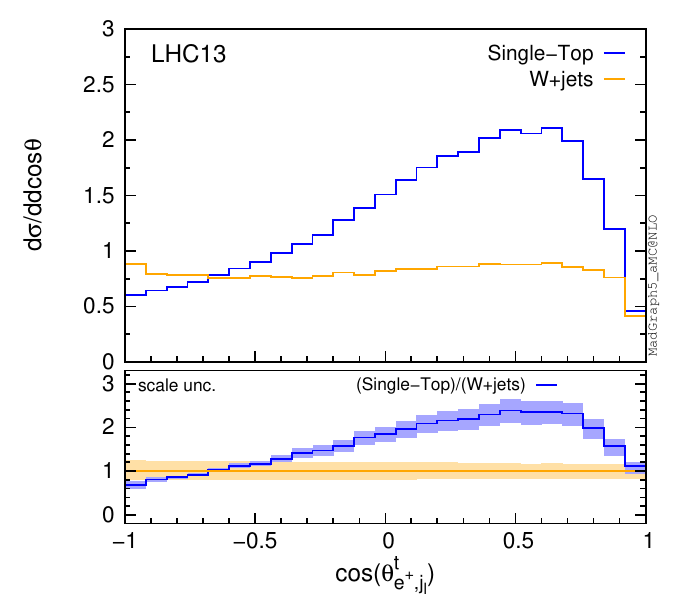}
\includegraphics[width=0.43\textwidth]{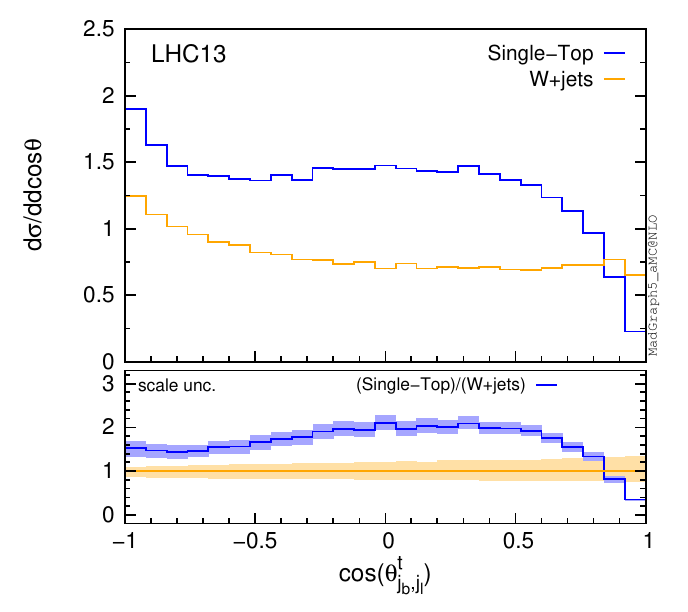}
\caption{Comparison between Single-Top and $W$+jets predictions at NLO QCD+EW accuracy for the same observables considered in Fig.~\ref{fig:FO2}.}

\label{fig:SB2}
\end{figure}

\medskip

Finally, in Figs.~\ref{fig:SB1} and  \ref{fig:SB2} we plot the predictions at NLO QCD+EW accuracy for Single-Top and $W$+jets for the same distributions considered in Figs.~\ref{fig:FO1} and \ref{fig:FO2}, respectively, and we show in the inset of each plot their ratio  {\SoB}, including scale uncertainties. In this case we do not normalise the $\coselj$ and $\cosbjlj$ distributions.  At the inclusive level {\SoB}$\sim1.7$, but several distributions show a very strong kinematic dependence for this quantity. 
 As expected, in the case of $\mttrue$ in the range shown in  Figs.~\ref{fig:SB2} the {\SoB}   ratio is much larger than at the inclusive level, however, this feature is strongly reduced for the experimental observable $\mtrec$. Non-negligible effects are also present for the remaining distributions. Notably, {\SoB} reduces to almost 1 for $\mebj$ in the region around 
the endpoint $\mebj=\sqrt{m_t^2-m_W^2}\sim 154 ~ {\rm GeV}$.

The case of $\eta(\ljet)$ is somehow special. As can be seen in Fig.~\ref{fig:SB1}, the ratio {\SoB} strongly increases moving from the central to the peripheral region.
This is particularly interesting for two reasons. First, the Single-Top $\eta(\ljet)$ distribution is rather flat, so cutting the central region the decrease of the total cross section is not dramatic. Second, as discussed in detail in Appendix \ref{sec:comparison}, the $tW$ contamination to the fiducial region we are considering is mainly affecting the central region of the $\eta(\ljet)$ distribution. Therefore, applying a veto on central light jets may at the same time improve the sensitivity to Single-Top production and also reduce the contamination from $tW$, leading to a measurement closer to the true $t$-channel single-top production. We have also verified that $t \bar t$ production, which is the main background in the measurements of $t$-channel single-top production via the signature \eqref{signature}, has much more central distribution for $\eta(\ljet)$ too.

The reader should note also that in Figs.~\ref{fig:SB1} and  \ref{fig:SB2}
scale uncertainties are associated to the NLO QCD+EW predictions, unlike in
the plots of Figs.~\ref{fig:FO1}-\ref{fig:FO3}, which display NLO QCD scale
uncertainties. As can be seen, in the case of Single-Top, the relative size of
the NLO QCD+EW scale uncertainties  is a bit larger than the NLO QCD ones in
specific phase-space regions. The main reason is that when NLO EW are large and negative, for example at large $p_T(\ljet)$ and $p_T(\bjet)$,  the central value is reduced and therefore the relative size of the scale uncertainties increases.

\subsection{Shower effects matched to NLO QCD}
In this section we provide numerical results at NLO QCD accuracy including shower effects for the signature \eqref{signature} within the cuts specified in Sec.~\ref{sec:signal}. Details on the calculational framework are described in Sec.~\ref{sec:showersetup}.  In Sec.~\ref{sec:totalshower} we present the inclusive results for separately $W$+jets and Single-Top production. In Sec.~\ref{sec:diffshower} we show and comment on differential distributions for Single-Top case only.
\label{sec:shower}

\subsubsection{Total cross section}
\label{sec:totalshower}

We start discussing the case of Single-Top production and then we move to the $W$+jets case.
In Tab.~\ref{table:FOvsPS_xsecs} we compare LO and NLO results for Single-Top
production including shower effects, denoted respectively as LOPS QCD and
NLOPS QCD, together with the corresponding results at fixed-order. In the same
table we also show three different ratios in order to separately display the
impact of shower effects and NLO corrections. As can be seen, the LOPS QCD
result is much smaller than the LO one, the shower effects reduce the cross
section by a factor 0.64. This effect is due to the jet veto, {\it i.e.},  the
request that there are at most two jets. We remind the reader that for this
reason results strongly depend on the $R^{\rm QCD}$ and $p^{\rm QCD}_{t, {\rm
    min}}$. On the other hand, once the shower effects are taken into account,
NLO effects have a no impact  at inclusive level with the ratio being at 1.00
and the NLOPS QCD result is higher by a factor of 1.08 w.r.t. the NLO QCD
one. At variance with the fixed-order case (see Tab.~\ref{table:FO}), NLO
effects do reduce the scale uncertainty, which moves form $_{-8.7 \%}^{+8.2
  \%}$ to $_{-3.3 \%}^{+3.3 \%}$. This result is important for two
reasons. First, it shows that if the actual experimental fiducial region is
considered, parton shower effects (or possibly analytic jet-veto resummation)
are necessary in order to reduce theory uncertainties for Single-Top
predictions. It is not even clear if NNLO QCD effects would reduce the scale
uncertainties or anyway would be useful for estimating the theory
error. Indeed, although in Ref.~\cite{Berger:2017zof} a reduction of scale
uncertainties is progressively observed moving from LO to NLO and then to NNLO
accuracy, this reduction is very sensitive to the definition of the fiducial
region (which is different in Ref.~\cite{Berger:2017zof} and in this work),
see also Appendix~\ref{sec:comparison}. Also, in Ref.~\cite{Berger:2017zof} it was pointed out that  LO, NLO and even NNLO QCD uncertainty bands do not overlap.   Second, NLO electroweak corrections at fixed order reduces the NLO QCD corrections by $\sim -3\%$, {\it i.e.}, their impact is at the same level with the scale uncertainty at NLOPS QCD. Summarising, both QCD shower and EW fixed-order effects are important in order to further improve the precision of predictions for the fiducial region.
\begin{table}[!t]
\begin{center}
\begin{tabular}{c c}
\toprule
Single-Top & cross section\\
\midrule
LO & $4.623(1)_{ -0.533(-11.5 \%)}^{+ 0.415(+9.0 \%)}$ pb\\
LOPS QCD  & $2.968(3)_{ -0.35(-11.9 \%)}^{+ 0.28(+9.3 \%)}$ pb\\
NLO QCD & $2.762(6)_{ -0.240(-8.7 \%)}^{+ 0.226(+8.2 \%)}$ pb \\
NLOPS QCD  & $2.974(9)_{ -0.098(-3.3 \%)}^{+ 0.098(+3.3 \%)}$ pb \\
(NLOPS QCD)/(LOPS QCD) &  1.00(1) \\ 
(LOPS QCD)/LO &  0.64(1) \\ 
(NLOPS QCD)/(NLO  QCD) & 1.08(1)\\ 
\bottomrule
\end{tabular}
\end{center}
\caption{Total cross section and scale uncertainty in various QCD approximations for the signature \eqref{signature} from Single-Top production within the fiducial region defined in Sec.~\ref{sec:signal}. The ratios are computed from the central values of their corresponding predictions.} 
\label{table:FOvsPS_xsecs} 
\end{table}
\begin{figure}[h]
\centering
\includegraphics[width=0.8\textwidth]{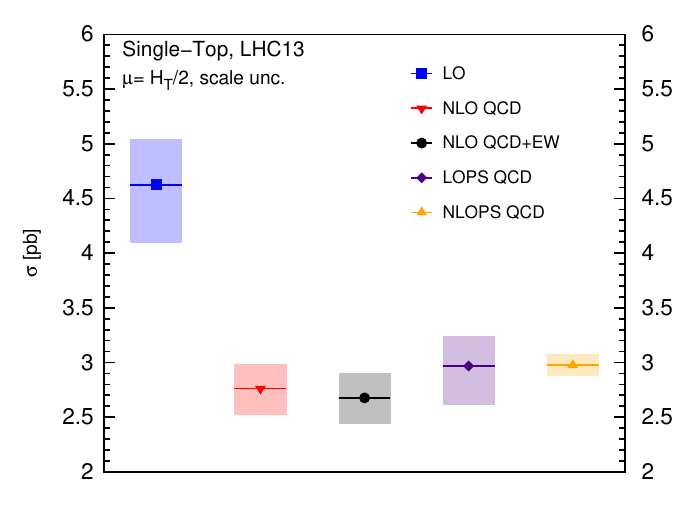}
\caption{Single-Top cross section and their uncertainty from scale dependence in the fiducial region in various approximations. Corresponding numbers are listed in Tabs.~\ref{table:FO} and \ref{table:FOvsPS_xsecs}.}
\label{fig:xs_N_LO_PS}
\end{figure}
However, these two results cannot be directly combined, being the latter based on a fixed-order computation for a process where shower effects are very large. The technology for matching NLO EW, and more in general complete-NLO, calculations to shower effects would be very useful for the calculation studied here.    We summarise in the plot in Fig.~\ref{fig:xs_N_LO_PS} the results obtained for the total cross section within the fiducial region, including the scale uncertainties, for different approximations discussed in this section and in Sec.~\ref{sec:total_fix}. 

\begin{figure}[!t]
\centering
\includegraphics[width=0.8\textwidth]{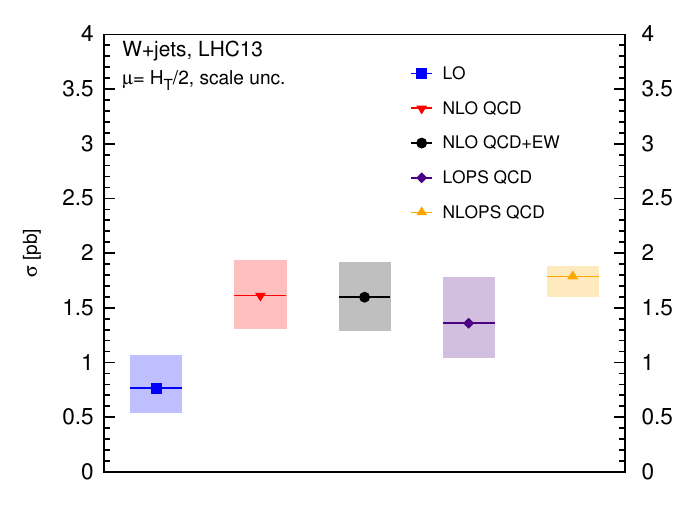}
\caption{$W$+jets cross sections and their uncertainties from scale dependence in the fiducial region in different approximations. Corresponding numbers are listed in Tabs.~\ref{table:FO} and \ref{table:FOvsPS_xsecsWj}.}
\label{fig:xs_N_LO_PS_Bckgrnd}
\end{figure}

Using the same layout of  Tab.~\ref{table:FOvsPS_xsecs} and Fig.~\ref{fig:xs_N_LO_PS} we show predictions at fixed order and including shower effects for $W$+jets production in  Tab.~\ref{table:FOvsPS_xsecsWj} and Fig.~\ref{fig:xs_N_LO_PS_Bckgrnd}.
At variance with the Single-Top case, the LOPS QCD result is larger than the
LO one; the shower effects increase the central value of the cross section by a factor 1.78. Although the jet veto (at most two jets) is present, the additional radiation induced by the shower splits part of the final-state gluons into $b \bar b$  pairs, converting events with two light jets into events with one $b$-jet and one light jet, which in turn contribute to the signature \eqref{signature}. 
\begin{table}[h]
\begin{center}
\begin{tabular}{c c}
\toprule
$W$+jets & cross section\\
\midrule
LO & $0.7656(6)_{ -0.2265(-29.6 \%)}^{+ 0.3002(+39.2 \%)}$ pb\\
LOPS QCD  & $1.36(2)_{ -0.32(-23.6 \%)}^{+ 0.42(+31.1 \%)}$ pb \\
NLO QCD & $1.612(3)_{ -0.309(-19.2 \%)}^{+ 0.323(+20.1 \%)}$ pb \\
NLOPS QCD & $1.79(5)_{ -0.18(-10.3 \%)}^{+ 0.09(+5.1 \%)}$ pb \\
(NLOPS QCD)/(LOPS QCD) &  1.31(4) \\ 
(LOPS QCD)/LO &  1.78(3) \\ 
(NLOPS QCD)/(NLO  QCD) & 1.11(3)\\ 
\bottomrule
\end{tabular}
\end{center}
\caption{Total cross sections and their uncertainty from scale
  dependence in various QCD approximations for the signature
  \eqref{signature} from $W$+jets within the fiducial region defined
  in Sec.~\ref{sec:signal}. The ratios are computed for the central
  values of the corresponding predictions.}
\label{table:FOvsPS_xsecsWj} 
\end{table}
Therefore, as explained in more details in Sec.~\ref{sec:total_fix}, the real
radiation leads to an increase of the cross section, even though a jet veto is
present.  Matching the shower simulation to NLO QCD corrections, and therefore
improving the simulation of hard real radiation, further increases the cross
section by a factor 1.31. In this case the central value of the NLOPS QCD result is higher by a factor of 1.11 w.r.t.~the NLO QCD one.

For $W$+jets the jet veto induced by the signature \eqref{signature} is not
leading to negative corrections, but nevertheless is preventing fixed-order
calculations to substantially improve the scale uncertainties moving from LO
to NLO. However, like in Single-Top,  taking into account shower effects, NLO
corrections do reduce the scale uncertainty, which moves from $_{-23.6
  \%}^{+31.1 \%}$ at LOPS to $_{-10.3 \%}^{+5.1 \%}$ at NLOPS QCD.  Therefore,
also for $W$+jets contributions to the signature \eqref{signature}, parton
shower effects (or possibly analytic jet-veto resummation) are necessary in
order to reduce theory uncertainties. On the other hand, the impact of NLO EW
corrections on top of NLO QCD predictions is much smaller ($\sim 1\%$ at the inclusive level) than the scale uncertainty even at NLOPS QCD accuracy. Given the results that we have found at the inclusive and differential level for NLO EW corrections (see Sec.~\ref{sec:fixres}), in the case of the $W$+jets contribution to the signature \eqref{signature}, the impact of EW corrections is negligible and their combination with showered effects is not so relevant as in the case of Single-Top.


\subsubsection{Differential distributions}
\label{sec:diffshower}

\begin{figure}[H]
\centering
\includegraphics[width=0.40\textwidth]{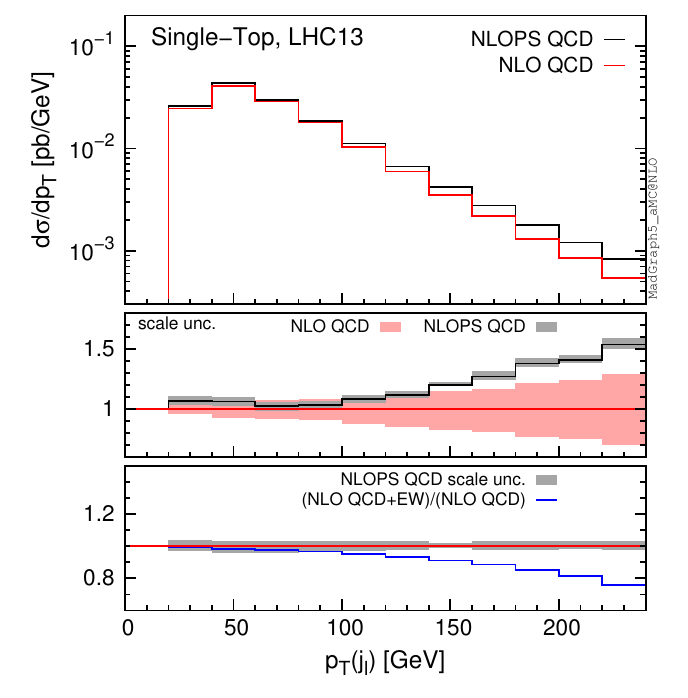}
\includegraphics[width=0.40\textwidth]{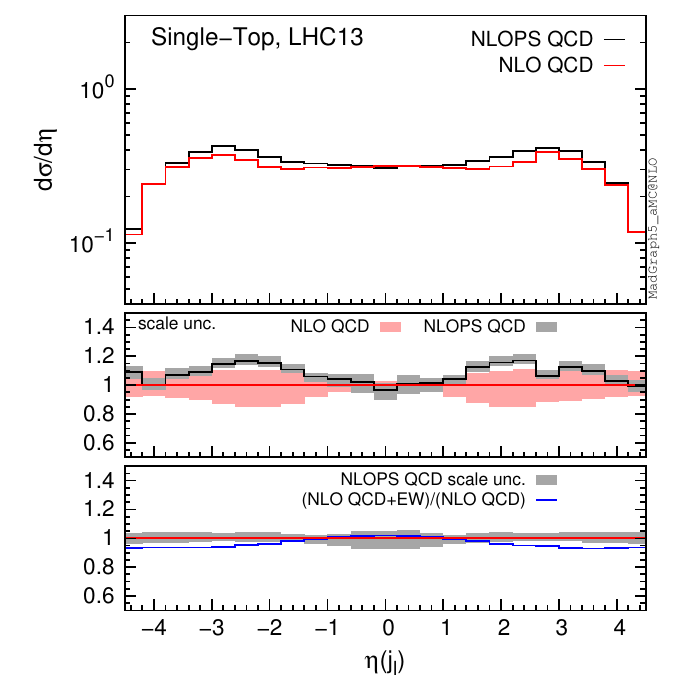}
\includegraphics[width=0.40\textwidth]{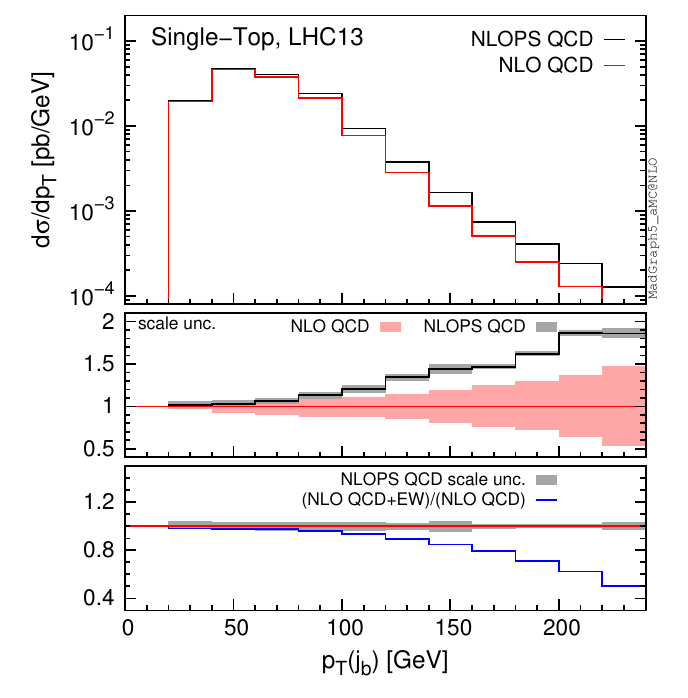}
\includegraphics[width=0.40\textwidth]{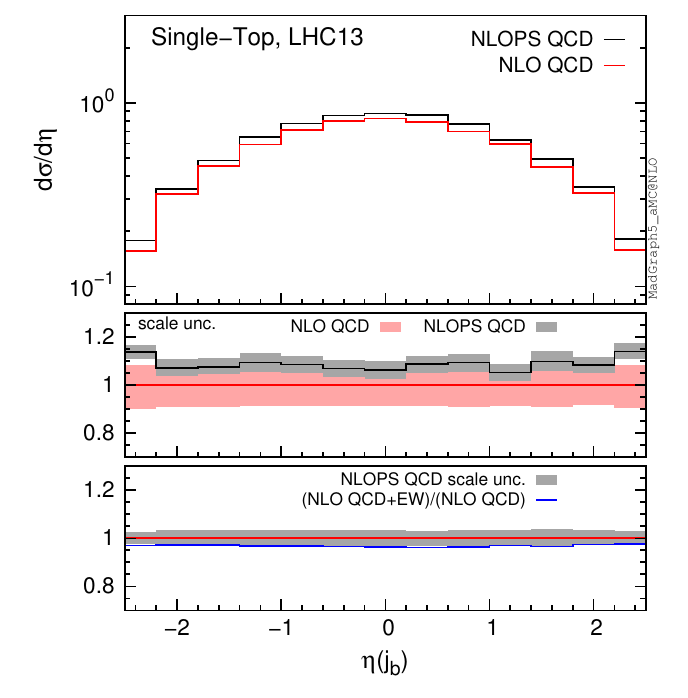}
\includegraphics[width=0.40\textwidth]{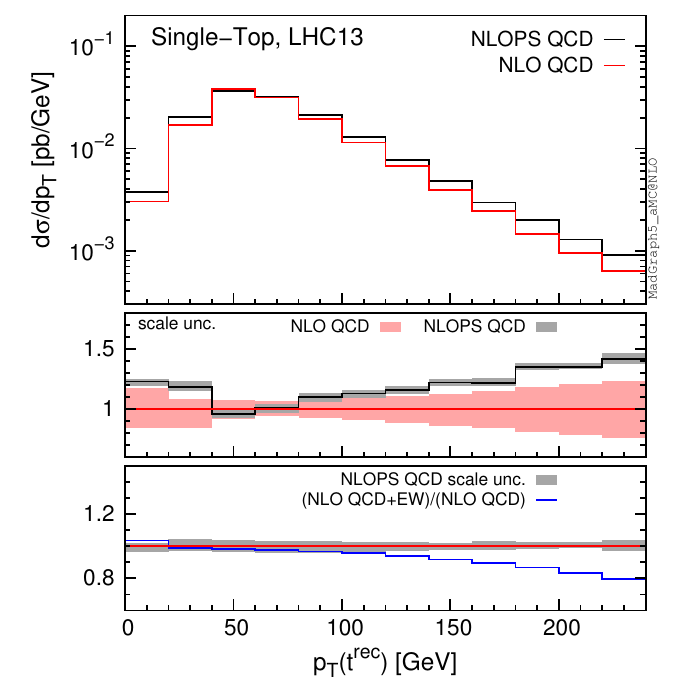}
\includegraphics[width=0.40\textwidth]{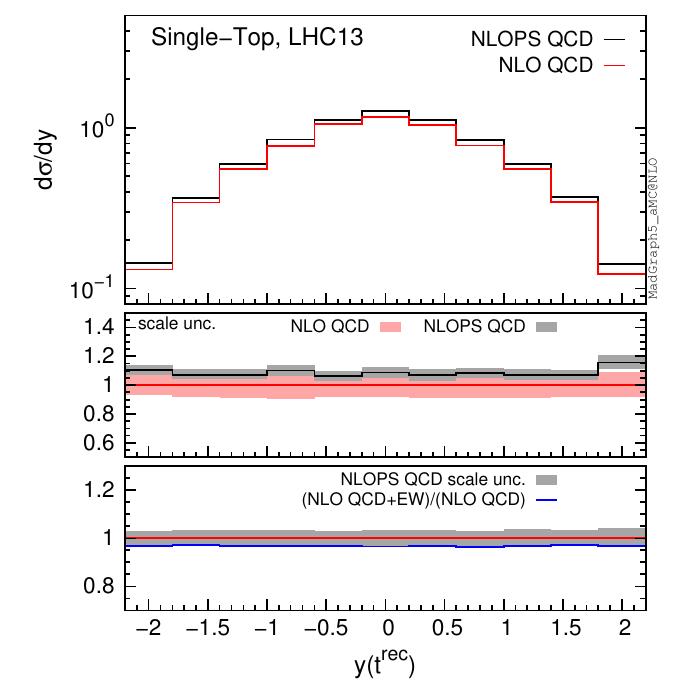}
\caption{Predictions at NLO(PS) QCD accuracy for the same observables
  considered in Fig.~\ref{fig:FO1} for the Single-Top process. Note
  that the second inset shows the ratio of the NLO QCD+EW over the NLO
  QCD process, but with the relative uncertainty from the NLOPS QCD
  superimposed on the latter.}
\label{fig:FO_PS_1}
\end{figure}

\begin{figure}[H]
\centering
\includegraphics[width=0.40\textwidth]{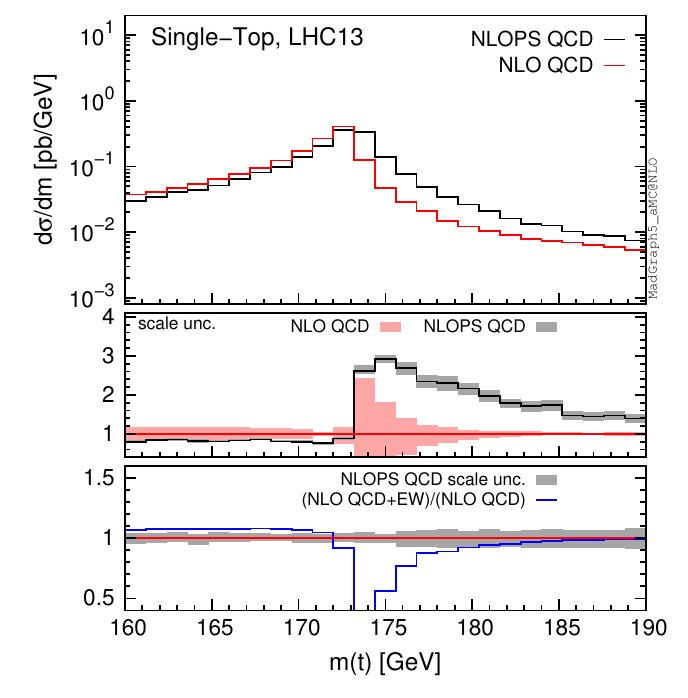}
\includegraphics[width=0.40\textwidth]{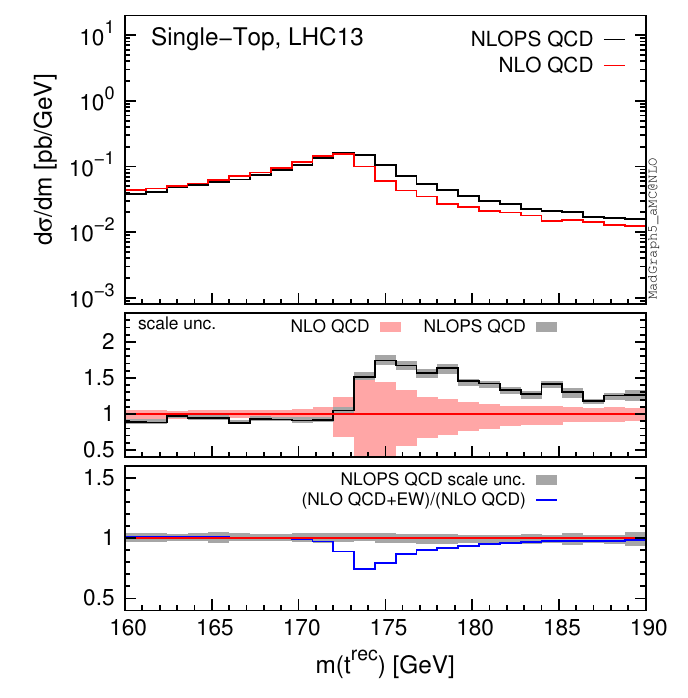}
\includegraphics[width=0.40\textwidth]{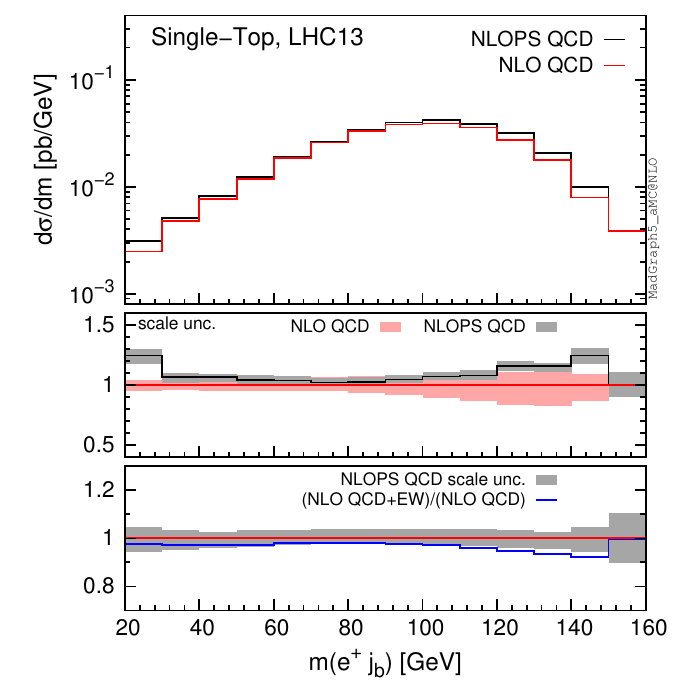}
\includegraphics[width=0.40\textwidth]{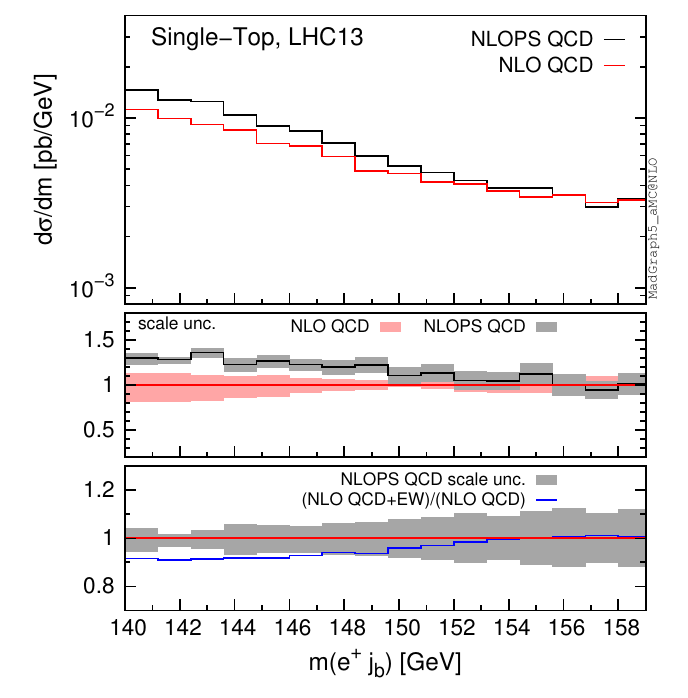}
\includegraphics[width=0.40\textwidth]{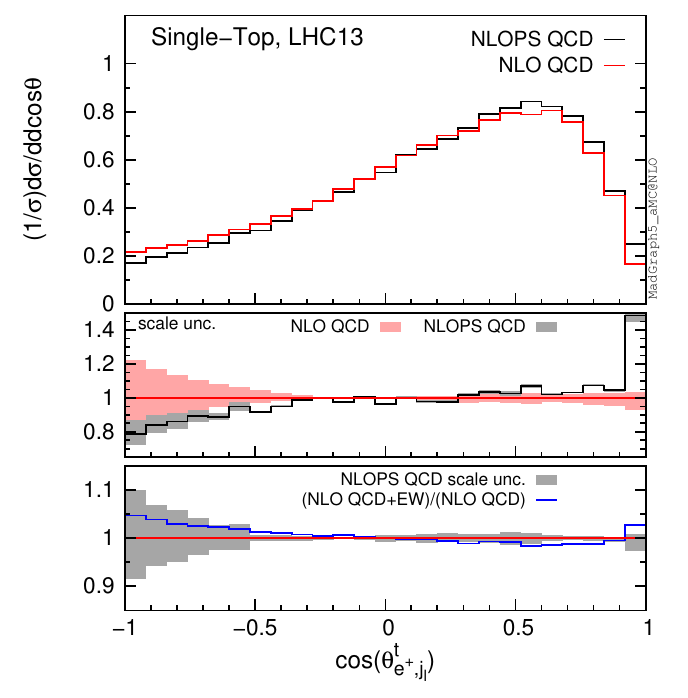}
\includegraphics[width=0.40\textwidth]{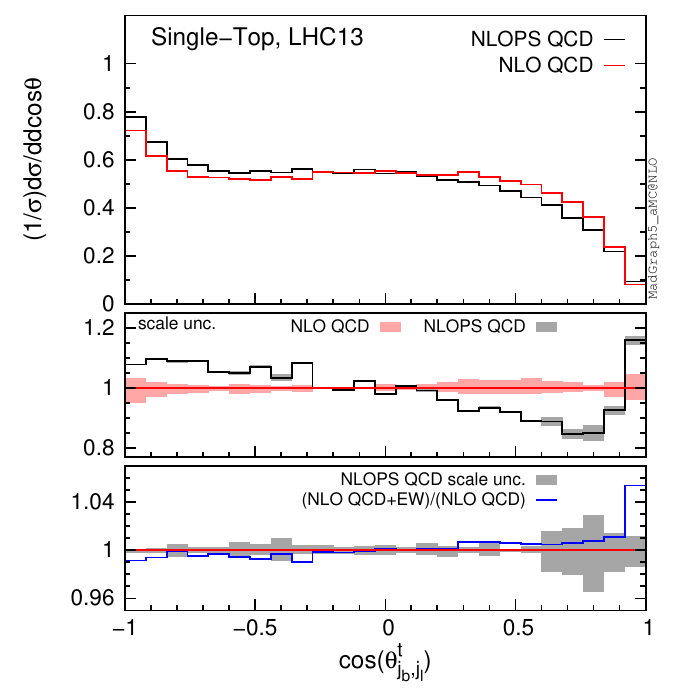}
\caption{Predictions at NLO(PS) QCD accuracy for the same observables
  considered in Fig.~\ref{fig:FO2} for the Single-Top process. Note
  that the second inset shows the ratio of the NLO QCD+EW over the NLO
  QCD process, but with the relative uncertainty from the NLOPS QCD
  superimposed on the latter.}
\label{fig:FO_PS_2}
\end{figure}

We move now to differential distributions for Single-Top production. In Figs.~\ref{fig:FO_PS_1} and \ref{fig:FO_PS_2} we show the NLOPS QCD predictions for the observables already considered in Figs.~\ref{fig:FO1} and \ref{fig:FO2}. In the main panel we show the central value for NLO QCD and NLOPS QCD predictions, while in the first inset we compare their scale uncertainties normalised to the central value of the NLO QCD prediction. Since the NLOPS QCD scale uncertainties are much smaller than the corresponding NLO QCD ones, it is interesting to compare them with the size of the NLO EW corrections, which on the other hand can be computed at the moment only at fixed order (see discussion in Sec.~\ref{sec:showersetup}). For this reason, in the second inset we show the NLOPS QCD scale uncertainty band normalised to its central value together with the  (NLO QCD+EW)/(NLO QCD) ratio already shown in Figs.~\ref{fig:FO_PS_1} and \ref{fig:FO_PS_2}. In other words, in such a way, we can directly compare the scale uncertainty at NLOPS QCD accuracy with the impact of NLO EW corrections on top of the corresponding fixed-order calculation. 


As at the inclusive level, NLOPS QCD scale uncertainties are in general
smaller than the NLO QCD ones and are roughly of the same size as the NLO EW
corrections in absolute value. However, shower effects are largely enhanced in
many regions of the phase space and especially the NLOPS QCD and NLO QCD predictions can be incompatible at the differential level. In the tails of the $p_T$ distributions for the light-jet, the $b$-jet and the reconstructed top-quark, scale-uncertainty bands are much smaller when shower effects are taken into account and especially they are clearly outside the corresponding NLO QCD scale-uncertainty bands. Moreover, in all the three cases, EW corrections are also much larger than the NLOPS QCD scale uncertainties. The corresponding (pseudo)rapidity distributions instead show compatible results at NLO QCD and NLOPS QCD accuracies, with anyway the latter with smaller scale uncertainties.  For these distributions, electroweak corrections are negative and, in absolute value, as large as the scale uncertainties at NLOPS QCD, with the exception of the central region for $\eta(\ljet)$.

As expected, the case of $\mttrue$ is extreme, especially for the region $\mttrue\gtrsim m_t$ where NLOPS QCD predictions are larger and far outside the NLO QCD ones. Also the region $\mttrue < m_t$ is strongly affected, with scale uncertainties very much reduced and decreasing the prediction to the lower edge of the NLO QCD scale-uncertainty band. In other words, the QCD parton shower has an opposite effect w.r.t.~NLO QCD corrections on top of LO, and therefore it flattens the distribution. Again, this distribution is strongly affected by the $R^{\rm QCD}$ and $p^{\rm QCD}_{t, {\rm min}}$ parameters. Around the peak, EW corrections are much larger than NLOPS QCD scale uncertainties and in minor extent also in the region $\mttrue < m_t$. However, similarly to the QCD case, the QED shower is expected to reduce these effects, which also strongly depend on the parameter $R^{\rm QED}$ for the clustering of photons with leptons. In the case of $\mtrec$ we observe the same features, but in a milder way. Similarly to the fixed order case, the realistic $\mtrec$ observable is flatter than the purely theoretical quantity $\mttrue$ and therefore migration effects from the peak are less severe.

The $\mebj$ distribution shows effects similar to those at the inclusive
level, with the exception of the phase-space region close to the end-point
$\mebj=\sqrt{m_t^2-m_W^2}\sim 154 ~ {\rm GeV}$, see central-left and -right
plots of Fig.~\ref{fig:FO_PS_2}. Indeed, in most of the phase space the
corrections from parton showering are small and positive, but significantly
reduce the NLO QCD scale uncertainties. However, close to the end-point, the
corrections increase to a maximum of $\sim 25 \%$ at $\mebj \simeq 140 ~ {\rm
  GeV}$; once we exceed the $\mebj\simeq 150 ~ {\rm GeV}$ and enter the
off-shell region the shower effects reduce very fast. Hence, the parton shower
partially compensates the large fixed-order NLO QCD corrections, and
introduces non-trivial alterations to the shape of the $\mebj$ distribution
close to the end-point.  A similar dynamics is also present for the region
$\coselj<-0.5$, where the shower effects reduce the NLO QCD prediction by up
to $20\%$. In the rest of this distribution, the shower effects are small,
except for the final bin, where the corrections are large and positive. In the
case of $\cosbjlj$, shower effects are small, similarly to the NLO corrections
at fixed order, with at most $\sim 10 \%$ effects. This observable therefore
turns out to be very stable under radiative corrections.

\section{Conclusions}
\label{sec:conclusions}

In this work we have calculated and provided precise predictions for the signature \eqref{signature}, which belongs to the class of those exploited at the LHC for the measurement of $t$-channel single-top production. All the results presented in the main text of the paper have been obtained by applying the cuts that are listed in Sec.~\ref{sec:signal}, defining the fiducial region.

First, we have calculated the complete-NLO predictions, {\it i.e.}, all NLO
effects of QCD and EW origin, for the signature \eqref{signature}. We have
shown that also other resonant processes contribute at this accuracy and we
provided predictions at the inclusive and differential level for all the
perturbative orders entering the complete-NLO results. According to the
underlying resonance structure (see Tab.~\ref{table:orders}) we have denoted
the orders $\LO_1$, $\LO_2$, $\NLO_1$ and $\NLO_2$ as $W$+jets and  $\LO_3$,
$\NLO_3$ and $\NLO_4$ as Single-Top (see also Eq.~\ref{conversion}). The
latter does include also contributions from the $s$-channel single-top,
$tW^-/\bar t W^+$ and $WZ$ production, which  we have not subtracted since we directly provide predictions for the signature \eqref{signature}.

Second, for both Single-Top and $W$+jets production we have calculated LO and NLO QCD predictions matched to parton shower effects. As discussed in Sec.~\ref{sec:showersetup}, the necessary technology for matching NLO QCD+EW predictions to shower effects, and possibly including also photon emissions in the shower, is not yet available. However our study clearly shows that it is desirable and, for particular observables, necessary in order to obtain precise and reliable predictions.

Here, we summarise our main findings.
At the inclusive level NLO EW corrections to Single-Top production are of order $-3\%$ w.r.t.~the NLO QCD prediction, which reduces the LO cross section by $\sim40\%$  and scale uncertainties (only) from  $\sim\pm10\%$  at LO to $\sim\pm8\%$ at NLO QCD. This effect is due to the presence of the jet-veto imposed by the signature \eqref{signature}. However, once parton-shower effects are taken into account, NLOPS QCD predictions reduce scale uncertainties to  $\sim\pm3\%$ and increase the fixed-order NLO QCD prediction by $8\%$. Notably, NLO EW corrections are in absolute value of the same order as QCD scale uncertainties when  parton-shower effects are included. We have also found that the impact of QCD corrections strongly depends on the cuts applied and in general on the definition of the fiducial region. Moreover, as documented in Appendix \ref{sec:comparison}, the contributions from $s$-channel and $tW^-/\bar t W^+$ production are sizeable also in the fiducial region.

At the differential level, both EW and QCD effects can be enormous and consequently also the effect from parton showering is very large. For instance, for values of $p_T \sim$ 250 GeV, for both the light- and $b$-jet, NLO QCD corrections reduce the LO prediction by $80\%$ and similar effects are present for the reconstructed top-quark mass close to the $m_t$ value. Moreover, in these phase-space regions, scale uncertainties are of order $\sim\pm50\%$ at NLO QCD, and even larger at NLO QCD+EW. However, shower effects reduce them to the order of  $\sim\pm3\%$, as at the inclusive level, and shift the predictions outside the scale-uncertainty band of NLO QCD predictions at fixed-order. The NLO EW corrections are also in general much larger than the percent level; in the case $p_T(\bjet)\sim250$ GeV they are of the order of $-50\%$ w.r.t.~the NLO QCD prediction. Therefore, since the origin of the enhancement is also in this case the  jet veto, the multiple emission of photons and QCD partons via EW interactions is also expected  to have a non-negligible impact in these specific phase-space regions. Also, NLO EW corrections are in general larger, or at least as large as, the NLOPS QCD scale uncertainty in absolute value over the full phase space. Last but not least, it is also important to remember that the opening of new resonating channels at NLO induces important distortions on distributions, such as the pseudorapidity of the light jet.

In the case of $W$+jets the situation is different. At the inclusive level, NLO EW corrections are of order $-1\%$ w.r.t.~the NLO QCD prediction, which increases the LO cross section by a factor $\sim2.1$ and reduce the scale uncertainty from  $\sim{}_{ -30 \%}^{+40 \%}$  at LO to $\sim\pm20\%$  at NLO QCD. The increase due to the NLO QCD corrections is quite surprising, given the presence of the  jet veto. However, as explained in the text, NLO QCD corrections induce $g\TO b\bar b$ splittings that lead to the migration of LO contributions inside the signature \eqref{signature}. Once parton-shower effects are taken into account, NLOPS QCD predictions reduce scale uncertainties to  $\sim{}_{ -10 \%}^{+5 \%}$ and increase the fixed-order NLO QCD prediction by $11\%$. Therefore, at variance with the Single-Top case, NLO EW corrections are in absolute value much smaller than QCD scale uncertainties also including parton-shower effects. 
At the differential level, both NLO QCD and EW corrections are mostly flat for the observable we considered. On the other hand, few exceptions are present and include, {\it e.g.}, the distribution of the rapidity of the light jet and the $p_T$ of the $b$-jet at small values.

 In conclusion, our study demonstrates  the relevance of both EW corrections
 and shower effects for obtaining precise and reliable theoretical predictions
 for the single-top-production fiducial region. Therefore, in this context,
 the possibility of performing NLO QCD+EW corrections matched with QCD and QED
 shower simulations would be desirable. Finally, from the experimental side,
 we  also suggest to study the possibility of applying a veto on central light
 jets in order to increase the Signal/Backgrounds ratio. Indeed such a cut suppresses much more the contributions of $W$+jets, $tW^-/\bar t W^+$ production and also $t \bar t$ production, which is the main background in the measurement of $t$-channel single top production, than the contribution from $t$-channel single-top itself.

\begin{acknowledgments}
This work has been supported in part by the Alexander von Humboldt Foundation, in the framework of the Sofja Kovalevskaja Award Project ``Event Simulation for the Large Hadron Collider at High Precision''. D.P.~has been also supported by the Deutsche Forschungsgemeinschaft (DFG) through the Collaborative Research Centre SFB1258. R.F.~is supported in part by the Swedish Research Council under contract number 2016-05996.

\end{acknowledgments}


\appendix

\section{Comparisons with previous results and different approximations}
\label{sec:comparison}

\begin{figure}[t]
\centering
\includegraphics[width=0.43\textwidth]{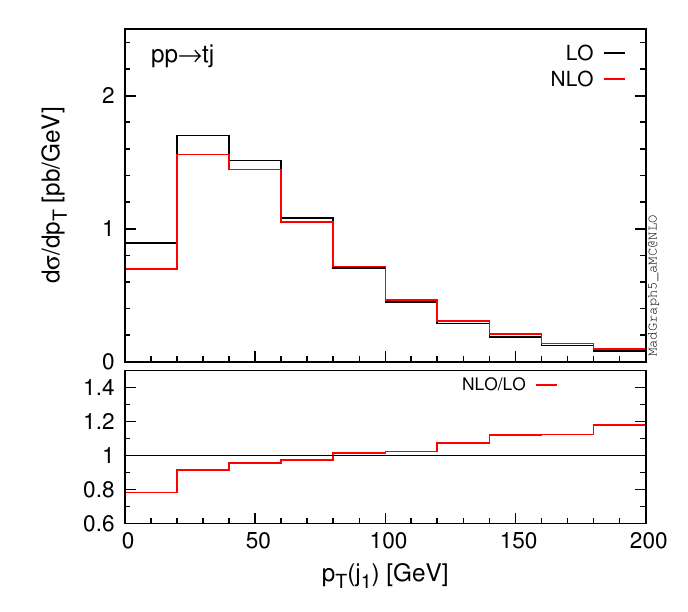}
\includegraphics[width=0.43\textwidth]{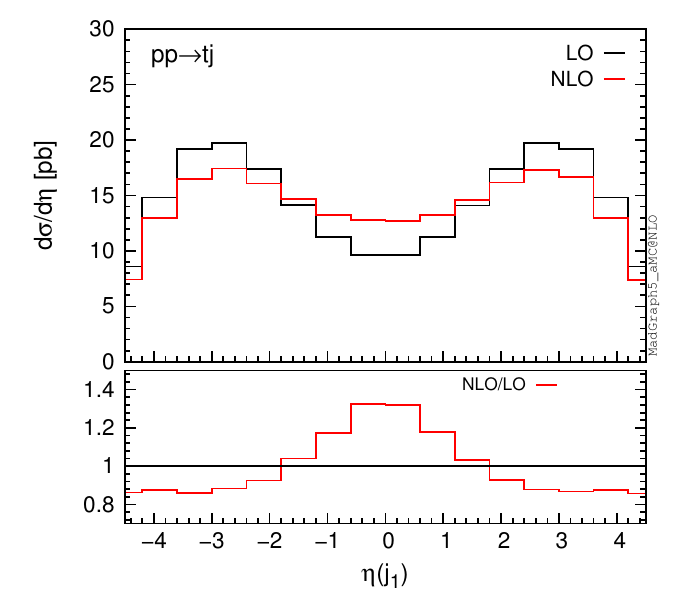}
\includegraphics[width=0.43\textwidth]{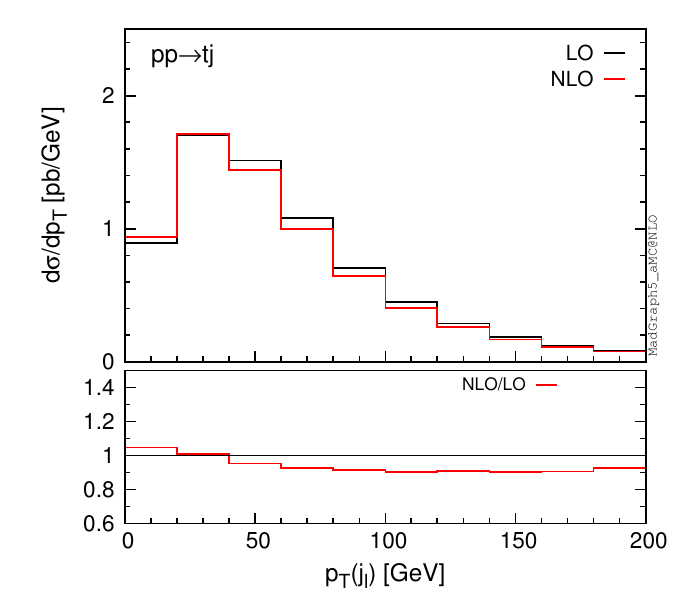}
\includegraphics[width=0.43\textwidth]{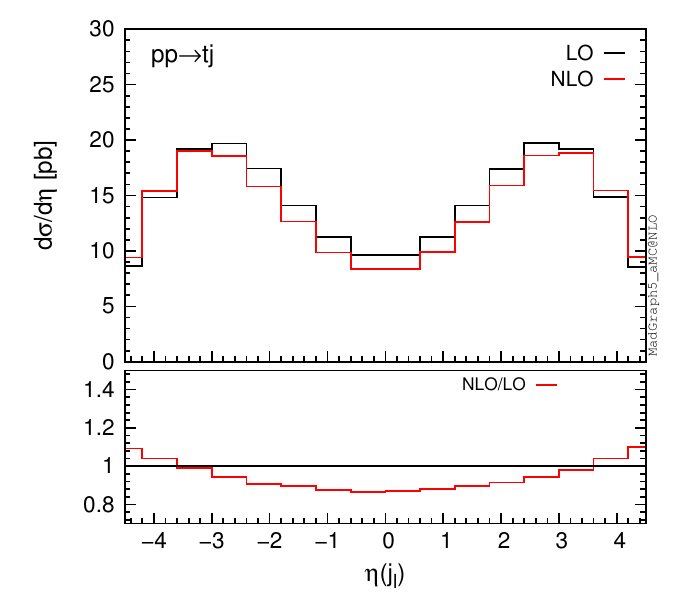}
\caption{Differential distributions for the process $pp\TO t\ljet$ at LO and NLO QCD accuracy. They can be compared with Figs.~20 and 22 (bottom-left plots) of Ref.~\cite{Berger:2017zof}.}
\label{fig:Prod_check}
\end{figure}

The purpose of this appendix is twofold. First, we want to compare our NLO QCD results for Single-Top production and decay with those presented in the literature, in particular the NLO results from Refs.~\cite{ Berger:2016oht} and \cite{Berger:2017zof}. These calculations used different input parameters\footnote{See Refs.~\cite{ Berger:2016oht} and \cite{Berger:2017zof} for the details.} and applied different cuts than those specified in Sec.~\ref{sec:signal}. Moreover, these calculations are performed employing a different approximation. Indeed, not only for the NNLO but also in the NLO case, only factorisable corrections in the narrow-width approximation for the $t$-channel $pp\TO t \ljet$ production with the subsequent leptonic top-quark decay are taken into account; non-factorisable corrections, non-resonant contributions, and the contributions from $s$-channel and $tW$ to the full final state have been ignored. Second, we want to explore the effects of the different cuts of the fiducial region for the signature \eqref{signature}. For this purpose we start from the case of $pp\TO t\ljet$ production and we progressively take into account the $t\TO Wb$ and $W\TO e^+ \nu_e$ decay and the effects of the cuts, directly comparing the obtained results with those of Refs.~\cite{ Berger:2016oht} and \cite{Berger:2017zof} for the same final state. We show results at the total cross-section level and for the light- and $b$-jets distributions. It is important to bear in mind that none of the aforementioned approximations are well-defined if we want to take into account EW corrections as done in Sec.~\ref{sec:fixres} ---here we are comparing only fixed-order QCD effects.

At the production level, $pp\TO t\ljet$, we find perfect agreement both at LO and NLO accuracy with the settings of Ref.~\cite{Berger:2017zof}, where no $p_T$ cut for the jet (clustered with $\Delta R_j=0.5$ and anti-$k_T$ algorithm) is set. Clearly, in order to obtain the agreement, we have excluded $s$-channel single-top contributions by excluding  any diagram with a $W$ boson in the $s$-channel. 
We also perform a comparison at the differential level for the leading jet, which accordingly to Ref.~\cite{Berger:2017zof} can be either a light or $b$-jet.

In the top plots of Fig.~\ref{fig:Prod_check} we show LO and NLO contributions
as well the $K$-factor for the transverse momentum and the rapidity of the
hardest jet.
Comparing these plots with the distributions in Figs.~20 and 22 of
Ref.~\cite{Berger:2017zof} we  find a very good agreement. The lower plots
show the case of the light jet (which is always identified with the hardest jet
only at LO). From the comparison of upper and lower plots we understand that $b$-jets emerging from real emissions are more central than the light jet.

We move now to the process $pp\TO W^+ b \ljet$. In this case, in order to be close to the calculation in Ref.~\cite{Berger:2017zof}, we use two different approximations. 
\begin{itemize}
\item Approximation A: we remove all diagrams with $W$, $Z$ and/or photon  $s$-channel propagators.
\item Approximation B: we remove all diagrams with $Z$ and/or photon (but not $W$) propagators.
\end{itemize}
In practice, the former does not include any single-top $s$-channel or $tW$ contribution, but it includes non-resonant effects with no intermediate top quarks already at LO. The latter instead involves only contributions from resonant diagrams at LO, which on the other hand includes also $s$-channel single-top contributions. Similarly, $tW$ contributions are present at NLO.   For the comparison with the results in  Ref.~\cite{Berger:2017zof} we calculate  the LO contribution from $pp\TO W^+ b j j $, with a $W^-$ and a top-quark in $s$-channel propagators, {\it i.e.}, the $tW$ contribution with top-quark leptonic and $W^-$ hadronic decays. Also, we calculate the LO contribution from $s$-channel single top ($pp\TO W^+ b \bar b $ vetoing any $Z$ or photon in the diagrams).  Since in the Approximation A initial-state collinear QED divergencies from $b\TO b\gamma$ splittings are already present at the LO, we use the generation cuts at the matrix-element level $\Delta R_j>0.5$, $p_T(j)>5$ GeV.

In Tab.~\ref{table:Int_xsecs} we display the numerical results for the calculations we have just described and we compare with the results from Ref.~\cite{Berger:2017zof}, which on the other hand are still at the purely production level $pp\TO t\ljet$. However, although we apply some technical cuts on the jets, we can still perform a qualitative comparison between results from Tab.~\ref{table:Int_xsecs} and Approximations A and B, being ${{\rm Br}(t\TO Wb)\simeq 1}$. More important is the use of a consistent value of $\Gamma_t$, which is very different at LO and NLO.

\begin{table}[t]
\small
\begin{center}
\begin{tabular}{c c c c c c c c}
\toprule
Order & Ref.~\cite{Berger:2017zof} [pb] & {\rm  A} [pb] & {\rm  B} [pb] & $tW$ [pb] &   $s$-channel [pb] \\
\midrule
LO  $(\Gamma_t^{\NLO})$& - & $157.88(1)^{+8.1 \%}_{-10.2 \%}$ & $163.96(2)^{+7.8 \%}_{-10.0 \%}$ & - & $5.15(2)^{+2.6 \%}_{-3.4 \%}$ \\
LO  $(\Gamma_t^{\LO})$& $144.5^{+8.1 \%}_{-10 \%}$ & $144.3(4)^{+8.1 \%}_{-10.3 \%}$ & $150.7(4)^{+7.8 \%}_{-9.9 \%}$ & - &  $4.73(2)^{+2.6 \%}_{-3.4 \%}$ \\
NLO QCD & $138.8^{+2.9 \%}_{-1.7 \%}$ & $137.8(3)^{+3.3 \%}_{-1.7 \%}$ & $160.5(1)^{+2.4 \%}_{-2.3 \%}$ & $19.1(1)^{+16.5 \%}_{-15.9 \%}$ & - & &  \\
\bottomrule
\end{tabular}
\end{center}
\caption{Comparison between the two approximations (A and B) described in the text, at the cross section level. The LO $tW$ process contributes to the NLO QCD in the approximation B.  At NLO we use $\Gamma_t^{\NLO}$.}  
\label{table:Int_xsecs}
\end{table}
The consistent use of $\Gamma_t$ is necessary for achieving the agreement at LO. Indeed, in Ref.~\cite{Berger:2017zof} the LO value of $\Gamma_t$ is used for LO calculations, while in the NLO calculations the  NLO value is used. Moreover, moving from the approximation A to B, there is a $\sim 5\%$ increase at LO due to the $s$-channel inclusion. At NLO QCD, also the $tW$ production is included and since at this point there are no cuts designed for $t$-channel single top, their effect cannot be ignored. At LO, the best agreement with the results of Ref.~\cite{Berger:2017zof} is achieved with the Approximation A, which does not include the $s$-channel. Similarly a good agreement is present for the value of the Approximation B subtracting the $s$-channel contribution. In both cases, the usage of $\Gamma_t^{\LO}$ is crucial. At the NLO level the Approximation A, but also the Approximation B subtracting the $s$-channel and $tW$ contributions have a qualitatively good agreement with Ref.~\cite{Berger:2017zof}.

\medskip 

\begin{figure}[!t]
\centering
\includegraphics[width=0.43\textwidth]{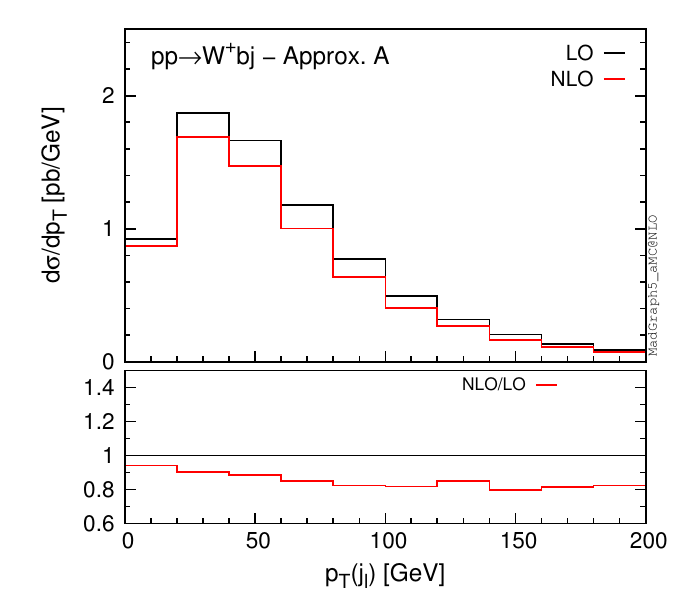}
\includegraphics[width=0.43\textwidth]{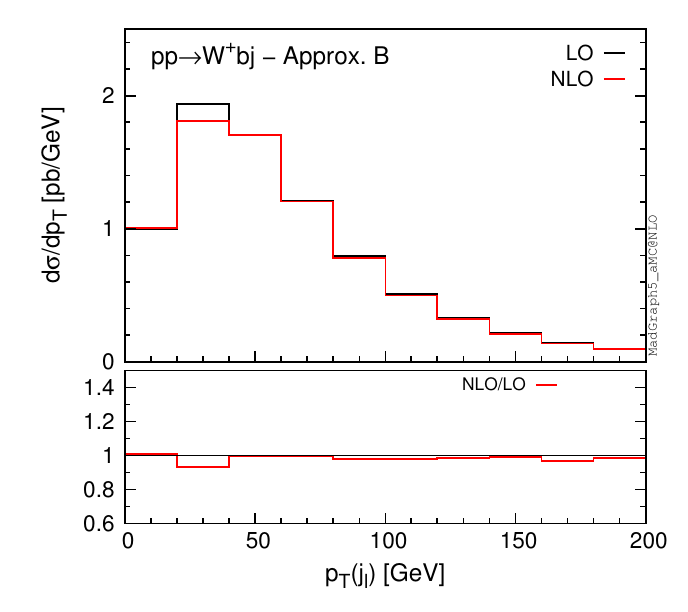}
\includegraphics[width=0.43\textwidth]{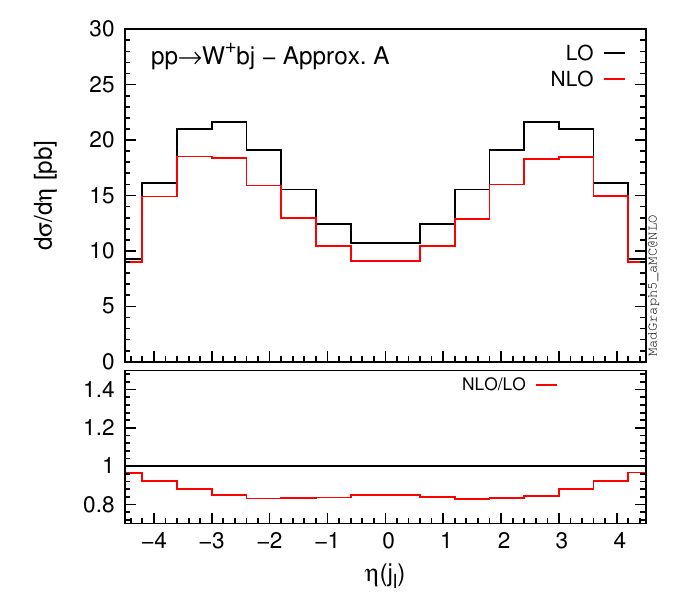}
\includegraphics[width=0.43\textwidth]{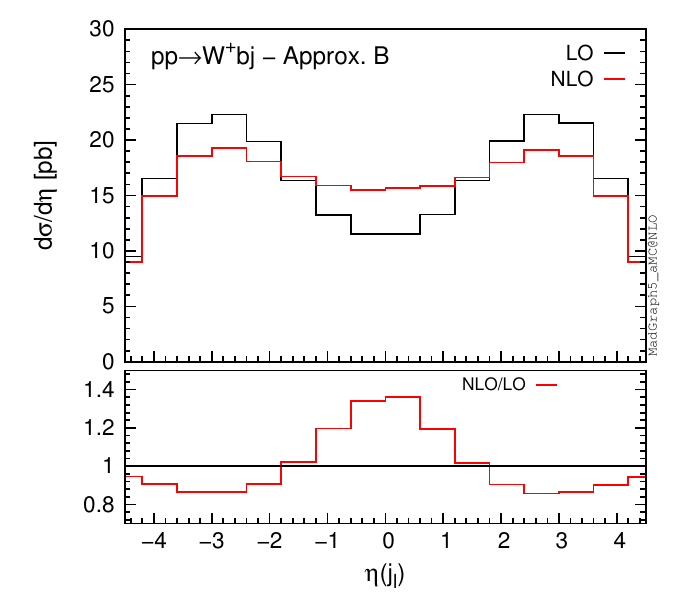}
\caption{The LO and NLO QCD predictions for the $p_T(\ljet)$ and $\eta(\ljet)$ distributions for the process $pp\TO t\ljet (t\rightarrow W b)$ using Approximations A (left) and B (right).}
\label{fig:Int_check_lj}
\end{figure}

We also show in Fig.~\ref{fig:Int_check_lj} a comparison between the
approximation A (left) and B (right) for  the $p_T$ (top) and the $\eta$
(bottom) distributions of the  (leading) light jet.
The inclusion of the $s$-channel single-top and the $tW$ production flattens the $K$-factor of the $p_T$ distribution and completely changes the shape of the $\eta$ distribution. The reason is that for both these two additional contributions the leading light jet is more central. So although we have considered only light jets, we see a similar effect to the one observed for the upper-right plot of Fig.~\ref{fig:Prod_check}

Then we compare Approximations A and B after applying part of the fiducial-region cuts of Sec.~\ref{sec:signal}. Specifically,
\begin{itemize}
\item light jet: $|\eta(\ljet)| < 5,~ \Delta R > 0.5,~ p_T(\ljet) > 40$ GeV,
\item $b$-jet: $|\eta(\bjet)| < 2.4,~ \Delta R > 0.5,~ p_T(\bjet) > 40$ GeV,
\item event veto: require exactly two jets of which exactly one is a $b$-jet.
\end{itemize}

 These cuts are expected to reduce the $s$-channel single-top and the $tW$ contributions. Especially, the last requirement suppresses the $s$-channel single-top, because this channel typically leads to two $b$-jets at LO, and the $tW$ because mostly leads to three jets
\begin{table}[!t]
\small
\begin{center}
\begin{tabular}{c c c c c c}
\toprule
 Vetoes & Order &  {\rm  A} [pb] & {\rm  B} [pb] & $tW$ [pb] &   $s$-channel [pb]  \\
\midrule
\multirow{2}{*}{No jet veto} & LO & $157.2(2)^{+8.1 \%}_{-10.3 \%}$ & $163.8(1)^{+7.8 \%}_{-10.0 \%}$ & - & $5.17(2)^{+2.6 \%}_{-3.4 \%}$ \\
& NLO QCD & $131(2)^{+3.5 \%}_{-2.7 \%}$ & $159(1)^{+2.4 \%}_{-2.2 \%}$ & $19.3(1)^{+16.6 \%}_{-15.9 \%}$ & \\
\multirow{2}{*}{$n_j=2,~ n_{\bjet}>1$} & LO & $74.9(1)^{+7.6 \%}_{-9.8 \%}$ & $79.2(2)^{+7.4 \%}_{-9.6 \%}$ & - & $2.83(2)^{+2.0 \%}_{-2.8 \%}$ \\
& NLO QCD & $47.7(6)^{+6.1 \%}_{-7.1 \%}$ & $59.3(6)^{+4.6 \%}_{-3.6 \%}$ & $9.28(8)^{+17.0 \%}_{-16.3 \%}$ & \\
\multirow{2}{*}{$n_j=2,~ n_{\bjet}=1$} & LO & $74.8(1)^{+7.6 \%}_{-9.8 \%}$ & $77.0(1)^{+7.3 \%}_{-9.4 \%}$ & - & $0.676(9)^{+0.9 \%}_{-1.6 \%}$ \\
& NLO QCD & $41.8(4)^{+8.4 \%}_{-5.7 \%}$ & $52.9(9)^{+6.0 \%}_{-4.0 \%}$ & $9.15(8)^{+17.1 \%}_{-16.3 \%}$ & \\
\bottomrule
\end{tabular}
\end{center}
\caption{Comparison between approximations A and B applying subsequent jet vetoes.}  
\label{table:Int_xsecs_fid}
\end{table}

In Tab.~\ref{table:Int_xsecs_fid} we show numerical results applying the aforementioned cuts without requirements on the number of ($b$-)jets and then progressively asking exactly two jets and at least one $b$-jet and then exactly one $b$-jet.
  The first row of the table actually differs with results of
  Tab.~\ref{table:Int_xsecs} only for the definition of the jets; the results
  are qualitative identical. On the contrary, both in the second and third
  line NLO corrections are negative and therefore reduce the value of the
  total cross section. Moreover, scale uncertainties are much larger at NLO as
  compared to no jet veto,  especially for Approximation A. This comparison clearly shows that vetoing extra jets leads to larger scale uncertainties. Also, these vetoes strongly suppress the $s$-channel contribution, but leave a non-negligible contribution from $tW$ production; it is $17\%$ of the results with Approximation B and accounts for most of the difference with the approximation A.
 
\begin{figure}[!t]
\centering
\includegraphics[width=0.43\textwidth]{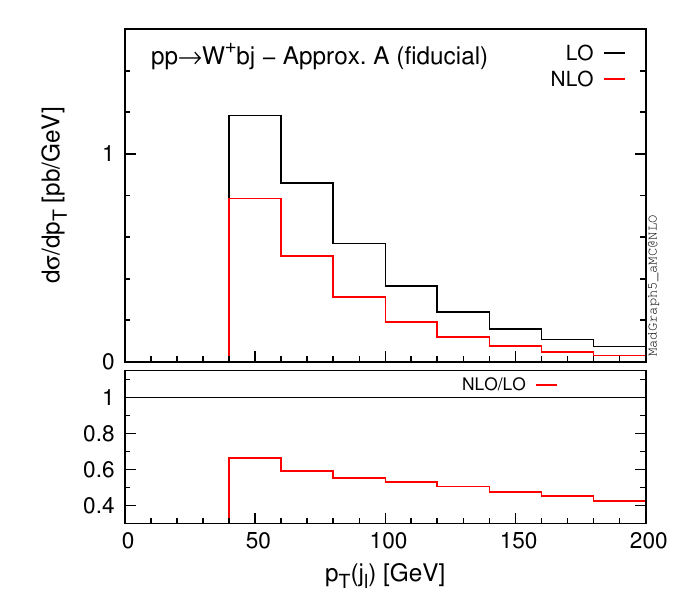}
\includegraphics[width=0.43\textwidth]{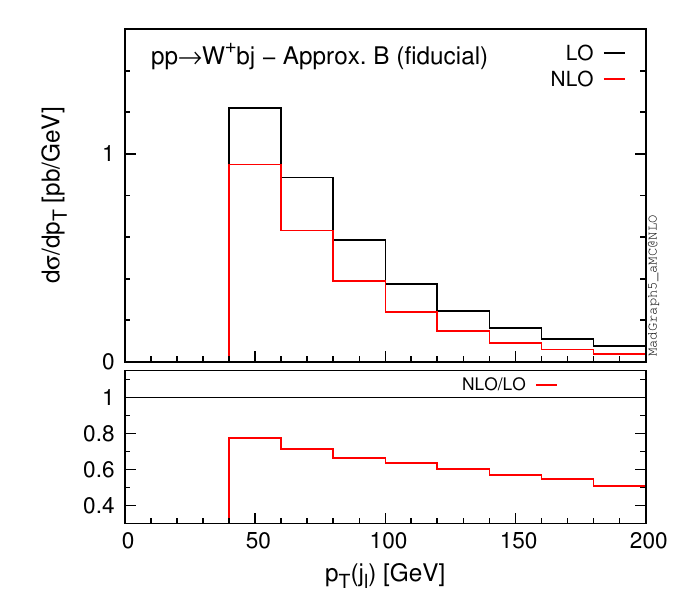}
\includegraphics[width=0.43\textwidth]{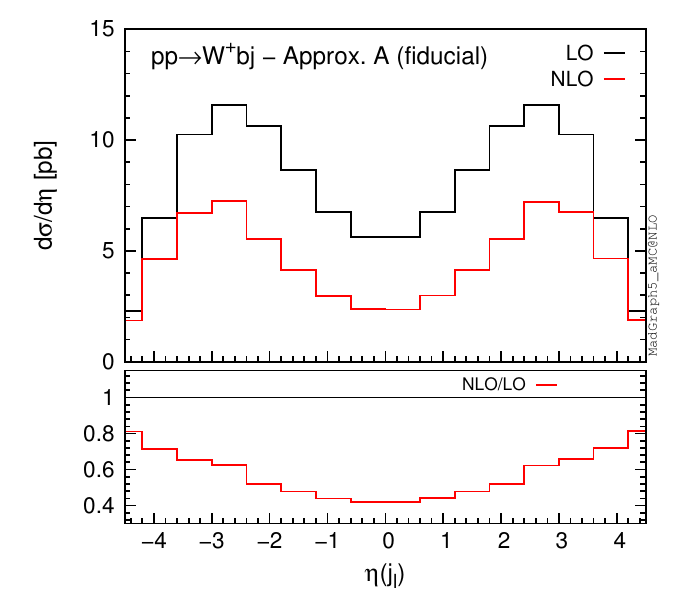}
\includegraphics[width=0.43\textwidth]{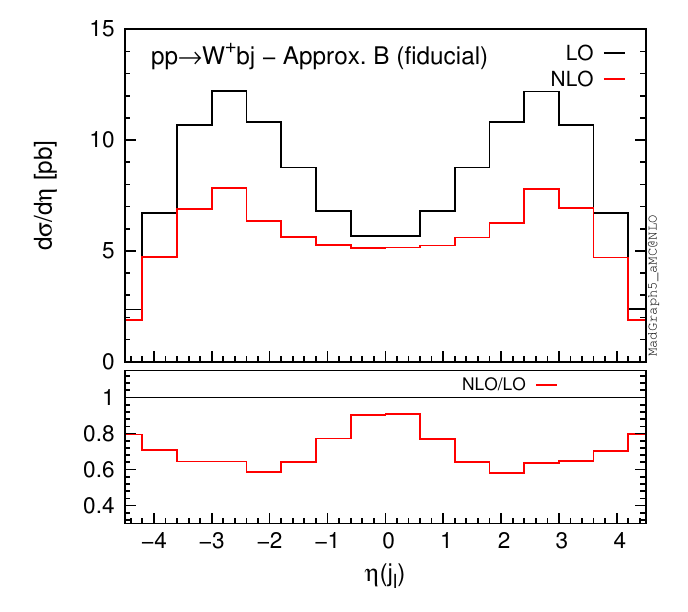}
\caption{Same as Fig.~\ref{fig:Int_check_lj}, but applying part of the fiducial-region cuts (see text for details).}
\label{fig:Int_check_lj_fid}
\end{figure}

We also show distributions corresponding to the case of the last row of  Tab.~\ref{table:Int_xsecs_fid}. 
 In Fig.~\ref{fig:Int_check_lj_fid} we show the $p_T$ (top plots) and $\eta$ (bottom plots) distributions of the leading light jet.
The shape of the $K$-factors for the $p_T$ is similar but this is not the case for the $\eta$ distribution. This implies that despite being in the fiducial region there are still some contributions with central leading light-jet, which clearly are not related to $t$-channel single-top production. Based on the results of Tab.~\ref{table:Int_xsecs_fid}, we can conclude that they mostly originate from the $tW$ production. 

Moving to the case where also the $W\TO e^+ \nu_e$ decay is included, we clearly cannot use the Approximation A. We compared the results obtained with the Approximation B with the results in Ref.~\cite{Berger:2017zof}, where decays and cuts on decay products are taken into account.
Taking into account that non-resonant effects are not present and that they use different scales for the production and the decay, we found again good agreement ($\sim 2\%$ difference) after subtracting the LO $s$-channel single-top contribution and the contribution from $tW$ production, which represents in total $\sim 20\%$ of the results in the Approximation B. At the differential level we observe the same features of Fig.~\ref{fig:Int_check_lj_fid}. Notably, similarly to Ref.~\cite{Berger:2017zof}, we observe also here a reduction of the scale uncertainties moving from LO to NLO, which instead we do not  see in the case of Single-Top in Tab.~\ref{table:FO}, where different cuts have been used. Therefore, the behaviour of the scale uncertainties is  very sensitive to the definition of the fiducial region. Moreover, in Ref.~\cite{Berger:2017zof} it was pointed out that  LO, NLO and even NNLO QCD uncertainty bands do not overlap. This comes with no surprise since we have shown in this paper that QCD shower effects are important. 

\pagebreak

\bibliographystyle{JHEP}
\bibliography{draft_bib}
\end{document}